\documentclass[useAMS,usenatbib]{mn2e}
\usepackage{txfonts, ulem}
\usepackage{graphicx}

\def\cm{\,{\rm cm}}

\def\ergscm2 {erg\,s$^{-1}$cm$^{-2}$}

\def\cm2 {cm$^{-2}$}

\def\aap {A\&A}
\def\apj {ApJ}
\def\prd {Phys. Rev. D}
\def\prc {Phys. Rev. C}
\def\mnras {MNRAS}
\def\pasj {PASJ}

 \title[Primordial dsQNe: VMP stars and their chemical composition]{A new model for the origin of very metal poor stars and their chemical composition}
\author[Rachid Ouyed]{Rachid Ouyed\thanks{email:rouyed@ucalgary.ca}\\ 
   Department of Physics and Astronomy, University of Calgary,2500 University Drive NW, Calgary, Alberta, T2N 1N4 Canada}

\begin{document}
\date{}

\pagerange{\pageref{firstpage}--\pageref{lastpage}} \pubyear{2012}

\maketitle

\label{firstpage}

\begin{abstract} 
 The genesis and chemical patterns  of the metal poor stars in the galactic halo remains an open question.
Current models do not seem to give a satisfactory explanation for  the observed abundances of Lithium in the galactic metal-poor stars  and the existence of carbon-enhanced metal-poor (CEMP)  and Nitrogen-enhanced metal-poor (NEMP) stars.  In order to deal with some of these theoretical issues, we suggest an alternative  explanation, where some of  the Pop. III SNe are followed by the detonation of their neutron stars (Quark-Novae; QNe). 
In QNe occurring a few days to a few weeks following  the preceding SN explosion, the neutron-rich relativistic QN ejecta leads 
 to  spallation of $^{56}$Ni processed in  the ejecta of the preceding SN explosion and thus to 
 ``iron/metal impoverishment"  of the primordial gas swept by  the combined SN$+$QN ejecta.
 We show that the  generation of stars formed from fragmentation of pristine clouds swept-up by the combined SN$+$QN ejecta   
 acquire a metallicity with $-7.5 < {\rm [Fe/H]} < -1.5$ for dual explosions   with  $ 2 < t_{\rm delay}\ ({\rm days}) < 30$. 
  Spallation leads to the depletion of $^{56}$Ni and formation of sub-Ni elements such as Ti, V, Cr, and Mn
  providing a reasonable account of the trends observed  in galactic halo metal-poor stars.
  CEMP stars form in dual explosions with short delays ($t_{\rm delay} < 5$ days). These lead 
    to important destruction of $^{56}$Ni (and thus to a drastic reduction of the amount of Fe  in the swept up cloud)
    while preserving the  carbon processed in the outer layers of the SN ejecta.
  Lithium is produced from the interaction of the   neutron-rich QN  ejecta with the outer (oxygen-rich) 
 layers of the SN ejecta.  A Lithium plateau with $2 < {\rm A(Li)} < 2.4$  can be produced in our model as well as a corresponding  $^6$Li  plateau with  $^6$Li/$^7$Li $< 0.3$. 
     \end{abstract}

\vfill\eject
\section {Introduction}

Galactic halo metal-poor  stars (i.e. with [Fe/H]$< -1$)\footnote{We adopt the notation [A/B]$\equiv \log(N_A/N_B)- \log(N_A/N_B)_{\odot}$,
where $N_A$ and $N_B$ refer to the number of atoms of elements $A$ and $B$, respectively. 
Elemental abundance can also be referred to as an ``absolute" scale, relative
 to the number of hydrogen atoms, defined by 
 $A(Li)=\log{(\epsilon(A))} =  \log{\frac{N_{\rm A}}{N_{\rm H}}} + 12$
 where $A$ can be taken to represent any element. The relevant  solar values are 
 taken from (Asplund et al. 2009; see their Table 1). The observed elemental abundances in stars are assumed to pertain
to their our atmospheres (where the lines are observed), not a volume average over their interiors.}
are are long-lived, low-mass objects ($0.6$-0.8$M_{\odot}$),
 the majority of which are main-sequence and
 giant stars that have preserved in their atmospheres the chemical signatures of the gas
  from which they formed.   It is argued that their stellar surface
    composition has not been significantly altered by any internal  mixing processes 
    or by external influences such as accretion of interstellar material.
  By measuring their surface composition today, one can look back in time and learn about
   the nature of the early Universe. 
      Thousands of very metal-poor  stars below [Fe/H]$\simeq -2$ have been identified in the Galactic halo by two large-scaled surveys (HK survey, Beers et al. 1992: Hamburg/ESO [HES] survey, Christlieb et al. 2001). The follow-up observations with high-dispersion spectroscopy (HDS) using large telescopes yield details of hundreds of these starÕs surface abundances.  Following the nomenclature defined in Table 1 in Beers\&Christlieb (2005; see also Frebel\& Norris 2011) which classify different types of metal-poor stars in terms of population, metallicity and chemical signatures, we refer to Metal-Poor stars as MPs (with [Fe/H]$<-1$), Very Metal-Poor stars as VMPs (with [Fe/H]$<-2$),
    Extremely Metal-Poor stars as EMPs (with [Fe/H]$<-3$), Ultra Metal-Poor stars as UMPs (with [Fe/H]$<-4$), Hyper Metal-Poor stars as HMPs (with [Fe/H]$<-5$) and, Mega-Poor stars as MMPs (with [Fe/H]$<-6$).
    The most iron-deficient objects known to date are the HMPs  identified among the HES sample
     with [Fe/H]$ = -5.6, -5.4$ and $-4.8$ (HE1327-2326, Frebel et al. 2005: HE0107-5240, Christlieb et al. 2002: HE0557-4840, Norris et al. 2007, respectively).
    
   Galactic halo MP stars  have received considerable attention in the literature because of the fact that they are at the crossroads of stellar evolution, star formation, galactic chemical evolution and cosmology. This has led to a 
      volumetric literature that has accumulated over the years providing 
   a thorough analysis  of the chemical abundances of MP
  stars and in particular on  their lithium (plateau) abundance (discussed in \S \ref{sec:introLi} below). 
   It is very difficult to give a complete and thorough introduction to the field of  MP stars and
   the issue of the Spite plateau.  The references in this paper  are most likely incomplete so 
  we refer the interested reader to the  reviews (and the numerous
  references therein) by  Lambert, (2004), Beers\&Christlieb (2005), Charbonnel \& Primas 2005), Sneden et al. (2010),  Aoki et al. (2010) and  more
  recent ones by  Karlsson, Bromm,   \& Bland-Hawthorn (2011); Cowan et al. (2011), Frebel\&Norris (2011) and Fields (2011). 
     We also suggest Iocco (2012) for a phenomenologist's perspective on the issue of the Lithium plateau.
     
  The conditions for the formation of these first low-mass stars are still being investigated.  
   Many formation and evolution models have been presented to tackle key questions
  relevant to MP stars and the plateau.
   Some studies propose that metal enrichment by first (massive) stars accelerates the cooling of molecular clouds to lower the Jeans mass and to enable the low-mass star formation  (Bromm\&Loeb 2003; Omukai et al. 2005). 
  Other studies propose that ionization photons emitted from first stars accelerate the formation of ${\rm H}_2$ molecules as main coolant in the primordial gas (Ricotti et al. 2002; Yoshida et al. 2008), which can trigger the first formation of low-mass stars (e.g. Shigeyama\& Tsujimoto 1998; Machida et al. 2005). 
  Any formation model of these stars should account for the abundance patterns of MP stars
   and in some case for unique patterns. E.g. all of three HMP stars, known to date, show large carbon enhancement (e.g. Beers\&Christlieb 2005).    
Others show enhancement of neutron-capture elements (e.g. Sneden et al. 1996).

\subsection{From metal poor to hyper metal-poor stars}

    Many attempts have been made to reproduce the overall abundance patterns in EMP and HMP stars by modelling the evolution and explosion of massive stars, and comparing the yields to observations (Woosley et al. 2002; Umeda\&Nomoto 2005; Nomoto et al. 2006;  Joggerst et al. 2010, to cite only a few). Some studies have focused on the potential role of very massive stars ($M \sim 200M_{\odot}$) that die as pair-instability SNe (PISNe) yielding  a large amount of metals  (Umeda\&Nomoto 2002; Heger\&Woosley 2002).  
    However, there seems to be no signatures of PISNe in  Galactic Halo EMP stars  (e.g. Ballero et al. 2006; Cayret et al. 2004; Tumlinson et al. 2004). 
    Other models  favour the idea that the observed EMP stars of the galactic halo are stars formed directly from the ejecta of one or few Pop. III SNe of intermediate mass (15-60$M_{\odot}$), which are diluted with metal-free primordial gas.  Yet other models   propose that HMP stars are formed of the gas, enriched with metals by the ejecta of peculiar faint Pop.~III SN
  (Umeda et al. 2003; Iwamoto et al. 2005).  
  Limongi et al. (2003)  propose  a combination of two Pop.~III SNe to reproduce the abundance pattern peculiar to HMP stars. 
These  scenarios are shown to face challenges  reproducing C, N, and O abundances simultaneously (e.g. Weiss et al. 2000
and references therein).
   
   Very energetic hypernova (HN) explosions  (driven by bipolar jets) may be responsible for these abundance patterns.  
    HNe are core-collapse SNe with explosion energies  up to an order of magnitude those of 
    normal core-collapse SNe (e.g. Iwamoto et al. 2003). These are known to show exceptionally
     large kinetic energies (Galama et al. 1998; Mazzali et al. 2000).   HNe are able to fit the average abundances of 
       VMP stars in particular those with no enhancement of CNO elements (e.g. Nomoto et al. 2006).
    The aspherical  yields of large (Co,Zn)/Fe and small (Mn,Cr)/Fe  are consistent with abundance patterns in MP halo stars, indicating important contribution of hypernovae in the early Galactic chemical evolution.
        On the other hand, Heger \& Woosley (2010) studied the evolution and parameterized explosions of Pop III stars with masses ranging from 10 to 100 $M_{\odot}$. They have concluded that the EMP stars do not show the need for an HN component and that moderate  explosion energies seem to be preferred, though their models tend to underproduce Co and Zn. Also, the explosion mechanism for HNe has yet to be identified.

     A  binary scenario was put forward to assert that HMP stars can be the survivors of the low-mass Pop.~III stars without pristine metal elements.  
   In this scenario, their surface metal elements originate from afterbirth pollution by the accretion of interstellar matter (ISM) and by the wind accretion of envelope matter ejected by the binary companions (e.g. Suda et al. 2004).   
   In this respect, Nishimura et al. (2009) investigated the AGB nucleosynthesis under the dearth of pristine metals to show that the abundance patterns of light metal elements from C through Al can be reproduced in terms of the accretion of envelope matter from AGB companion in binaries.  
  However,  in this case, the origin of the iron-group elements remains to be discussed because binary mass transfer or stellar wind cannot provide these elements.

    A large fraction of VMPs show a great enhancement in the abundance of C.
   These carbon-enhanced MP (CEMP) stars constitute approximately one fifth of the MP ([Fe/H] $<-2$) population (Beers\&Christlieb 2005). 
   Apart from carbon, substantial enhancements of N with respect to iron are common among CEMP stars. One explanation of this observation is that these stars have undergone mass transfer from an AGB binary companion.     
         The observed [C/N] ratios in CEMP stars do not fit the predictions of either the low-mass or the intermediate-mass AGB
    models but instead fall in an intermediate regime that is not covered by the models.
     This begs the question of why do the observations of stars with large [C/Fe] show larger N (smaller [C/ N]) than predicted by models for the evolution of 
     2-3$M_{\odot}$ stars?  One also wonders where are the VMP stars that were, in the context of the mass transfer scenario, polluted by the N-rich 3.5-7.5$M_{\odot}$ stars? 
       The mechanism responsible for the very efficient production of N in intermediate-mass stars is a robust prediction of stellar evolution models. If the binary mass transfer scenario is invoked for the CEMP-s stars with [C/N]$\sim$1 (CEMP-s stars are CEMP stars that exhibit enhancement of the neutron-capture s-process elements), then one may ask why there are no NEMP stars with [C/N]$\sim$ -1, as might be expected to arise from systems in which the donor is a more massive AGB star. 
        A possible  explanation for both the ubiquity of CEMP-s stars and the near-absence of NEMP stars is that low-mass, low-metallicity AGB stars undergo much more efficient dredge-up of carbon than shown by detailed evolution models available to date (see Izzard et al. 2009).
     This   has inspired renewed interest in the nucleosynthesis of N in massive stars, 
  which has led to the suggestion that N  in EMP and HMP stars may require rotationally induced mixing prior to the destruction of the star
   (Chiappini et al., 2005; Hirschi, 2007; Ekstr\"om et al. 2008; Joggerst et al. 2010). 
   The idea that  Pop III stars were  rotating\footnote{The non-rotating Pop III model yields achieved only partial agreement with the abundances in the most MP stars ever detected. In particular, they did not reproduce the amount of nitrogen observed in HMP stars because it was not present in the initial pre-supernova progenitor models in sufficient quantities (e.g. Joggerst et al. 2010 and references therein).} needs to be confirmed. Furthermore, 
 while  the N  yield from these models does fit the  abundance pattern of a few halo stars,  the  C abundance was not easily reproduced.

\subsection{The Lithium Plateau}
\label{sec:introLi}

From observations of 11 main-sequence (Pop.II) stars belonging to the Galactic halo, Spite \& Spite 
(1982)
concluded that  the abundance of $^7$Li in halo stars is almost independent
 both of their effective temperatures (between 5600 K and 6400 K) and of their metallicity (between [Fe/H]$ = -1.5$ and -2.5).
Three decades of  work followed, increasing the accuracy, the number of stars observed and the range of metallicity that they span,
 which confirmed the plateau with $A(Li)=\log\epsilon_{\rm Li}\simeq 2.15\pm 0.1$ for $-3.5 < {\rm [Fe/H]} < -1.5$
 (Hobbs \& Duncan 1987; Rebolo et al. 1988; Spite \& Spite 1993; Thorburn 1994; Molaro et al. 1995; 
 Ryan et al. 1996; Bonifacio \& Molaro 1997;  Bonifacio et al. 2002; and Asplund et al. 2006). These studies found that 
the star-to-star	scatter in Li abundance for MP stars on the plateau is extremely small, 
 on the order of 0.05 dex, well within the expected observational errors (Ryan et al. 1999;
 Asplund et al. 2006; Bonifacio et al. 2007, to cite only a few).
 
One way to understand the apparently constant $^7$Li abundance on the  Spite
 plateau is to relate it directly to production by big bang nucleosynthesis (BBN). 
The measurements of the WMAP satellite allowed to determine the parameters of the model of nucleosynthesis of the BB, leading to a measured lithium abundance  about 3 times  (i.e. 0.5 dex) higher than the value of the  
 Spite plateau (Coc et al. 2004). Several theories proposed uniform partial destruction of the (high) big bang lithium, forming in this way a (low) plateau in agreement with observations.   Some investigated models involving non-standard models of the BB,  and more elaborate 
  nuclear physics in the early universe. Other ways in which to remove the two discrepancies involve aspects of a new cosmology, particularly through the introduction of exotic particles (see Lambert 2004).  However the discrepancies seems to resist explanation.

  Is there a probable stellar solution to the cosmological lithium discrepancy?
 Lithium is known to be a very fragile element, one that is rapidly destroyed by nuclear reactions in stellar interiors when the temperature exceeds $2.5\times 10^6$ K. One possible explanation of the discrepancy between the expected primordial lithium and that measured on the Spite plateau could therefore be that the lithium currently observed in the outer atmospheres of halo stars has been depleted by stellar-evolution processes over the long history of the Galaxy.  It is understandable that for some  it remains hard to admit  that the abundance of  lithium in the atmosphere of the old VMP dwarfs  had remained unchanged during 13 billions years. The challenge this school of thought faced was in explaining the strictly
          uniform (i.e. independent of temperature, mass and metallicity) lithium depletion in the warm MP dwarfs. 
          Richard et al. (2005) showed that the addition of turbulent mixing allows for a constant 
       lithium abundance with stellar mass, at the same time lowering the initial lithium by up to 0.5 dex thus favouring 
        the BBN origin.  In this respect, Korn et al. (2006; see also Korn et al. 2008)  find that both lithium and iron have settled out of the atmospheres of old stars in the MP globular cluster NGC 6397, and they infer for the un-evolved abundances, [Fe/H]$ = -2.1$ and  A(Li) = $2.54 \pm 0.1$, in excellent agreement with standard BBN.   This  result is facing scrutiny with recent investigations seemingly  disagreeing with these
        high values of $A(Li)$ and confirming the discrepancy 
     with BBN (e.g.  Bonifacio et al. 2007; Mucciarelli et al. 2012 and references therein). This is  a very complex problem that may require  much more 
     understanding of stellar astrophysics to be solved. 
Many efforts have been directed towards the investigation of the stellar solution but it appears that the conclusions
  are not firm because 
 low-mass stars exhibit, at all stages of their evolution, the signatures of complex physical processes that require challenging modelling beyond standard stellar theory (e.g. Charbonnel \& Primas 2005). 
 
  The $^6$Li observations of Asplund et al. (2006) hint at a plateau with  $^6$Li/$^7$Li$\sim 0.06\pm 0.03$ (or $\log{\epsilon (^6Li)}=0.8$;
these might be considered as upper values).  Concern was raised about the quality of the high-resolution spectra that have been used to determine $^6$Li/$^7$Li.   For example, line asymmetries generated by convective Doppler shifts in the atmosphere of metal-poor stars result in an excess absorption in the red wing of the $^7$Li absorption feature that mimics the presence of $^6$Li (see Cayrel et al. 2007). Neglecting such asymmetries in abundance analyses could lead to 
to overestimation of the $^6$Li/$^7$Li  ratio. However,  as emphasized by Asplund et al. (2006), one may question a detection of $^6$Li for a given star, but the collective distribution of  $^6$Li/$^7$Li can only be explained if $^6$Li is present in the atmospheres of some of these MP stars.   If confirmed,  the $^6$Li plateau (which still lacks convincing theoretical explanation) would widen the divide between the BBN lithium and
  the MP stars lithium. It would also call for a re-thinking of stellar physics and models of cosmic-ray nucleosynthesis. For example, the production of such large amounts of $^6$Li must have required an enormous flux of cosmic rays early in the history of our galaxy, possibly more than could have been provided by known acceleration mechanisms. Moreover, if the plateau stars have truly destroyed enough $^7$Li to bring the WMAP prediction of the mean baryon density into agreement with that obtained with the observed Spite plateau, the greater fragility of $^6$Li implies that the stars initially contained $^6$Li in quantities comparable to the observed $^7$Li plateau. 
 
Rollinde, Vangioni \& Olive (2006) have considered pre-galactic cosmic rays (CR) generated by shocks around population III objects. By collisions between accelerated and at rest helium nuclei, the so called $\alpha$ fusion process generates both Lithium isotopes by the reactions $^4He(\alpha, D)^{6}Li$ and $^4$He($\alpha$, p)$^7$Li. This process works, but needs a very large fraction of the kinetic energy of population III SNe to be converted into CR energy, as noted by Prantzos (2006) and Evoli et al. (2008). It seems also difficult to produce a plateau  in the hierarchical model of Galaxy formation of Evoli et al. (2008). Other scenarios involving CR production have been proposed, but are exposed to similar critics.
  The challenge has been since  to explain: (i) The source and the value of the plateau; 
(ii) the mere 0.05 dex after dispersion from one star to another;
(iii)  the gap (i.e. 0.5 dex) between the primordial Li and the Spite one? 
We refer to a recent paper by Spite\&Spite (2010) which discusses the
   challenges and the complexity inherent to this old enigma.
   To summarize,    replicating abundances and trends in Galactic halo MP stars have proven challenging.
  This leaves room for new alternatives to be explored.

\subsection{Our model}
  
  Here we present a model in the context of primordial Quark-Novae (hereafter QNe);  i.e. QNe going off in the wake of Pop. III stars. 
   A QN is an explosive transition of a massive neutron star to a quark star (Ouyed et al. 2002;
   Ker\"anen et al. 2005; Vogt et al. 2004; Ouyed et al. 2005; Niebergal et al. 2010).   
   Our previous studies of QNe favour progenitor initial mass in the 20-40 $M_{\odot}$; these progenitors (effectively
   the  SNe progenitors) would lead to the most massive NSs which are unstable to quark deconfinement
    in their cores (Staff et al. 2006) and prone to a subsequent QN transition/explosion (Niebergal et al. 2010). 
   The  20-40 $M_{\odot}$  mass range is
   in line with the literature which suggests prompt black hole (BH) formation above 40$M_{\odot}$ (e.g., Heger et al. 2003; Nakazato et al. 2008);
    the  most massive stars ($M > 40 M_{\odot}$) in a conventional or  top-heavy IMF collapse to black holes 
 while progenitors in the 20-40 $M_{\odot}$ range lead to QNe. 
   In the context of Pop. III stars,
  QNe mass range corresponds to the  lower end of Pop. III IMF  which implies that the dying out of the first heavy stars should coincide with a peak in the QN rate (Ouyed et al. 2009a).

   In particular QNe going off shortly (i.e. a few days to a few weeks) following the preceding/parent SN,
  lead to  spallation and  destruction of $^{56}$Ni in the SN envelope while producing sub-Ni elements (Ouyed et al. 2011).
  $^{56}$Ni spallation reduces  the subsequent Fe content in the SN ejecta following
   the decay of the leftover Cobalt.   This means that the cloud of pristine gas swept up
    by the SN+QN ejecta will carry signatures of  
    this  ``nickel/iron  impoverishment"   which should show up  in the next generations of stars formed from
     fragmentations of the swept up clouds. 
      As we show in this work, dsQNe in the wake of Pop. III stars provide 
        a universal origin of MP stars and of the Spite Plateau and seems to account for many of the
        unusual trends  observed in MP stars. 
 Before we describe   our model in details, we remind the reader of the QN model and its basic features.
   
 \subsubsection{Primordial Quark-Novae}
 
 The basic picture  of the QN is that a massive NS  converts 
explosively to a quark star (Ouyed et al. 2002; Ker\"anen et al. 2005).
Such an explosion can happen if the massive NS, in its spin-down evolution or via mass accretion (e.g.
from fall-back material), reaches the quark deconfinement density in its core  (Staff et al. 2006) and subsequently undergoes a phase transition to the conjectured more stable strange quark matter phase (Itoh 1970; Bodmer 1971; Witten 1984; see also Terazawa 1979), resulting in a conversion front that propagates toward the surface in the detonative regime
(Niebergal et al. 2010). The outcome of the QN explosion --  besides the formation of a quark star --  is the ejection of  the NS's outermost layers (a very neutron-rich
ejecta with an average mass $M_{\rm QN}\sim 10^{-3}M_{\odot}$) at  relativistic speeds (with a Lorentz factor averaging  $\Gamma_{\rm QN}\sim 10$; Ouyed\&Leahy 2009).   The outer layers  are ejected  from an expanding thermal fireball (Vogt et al. 2004;
Ouyed et al. 2005)  which  allows for ejecta with kinetic energy, $E_{\rm QN}^{\rm KE} \ge 10^{52}$ erg.
In previous papers, we introduced the QN as a model for superluminous SNe (Leahy\&Ouyed 2008; see also Ouyed\&Leahy 2012), discussed 
 their photometric/spectroscopic  signatures (Ouyed et al. 2012) as well as their nuclear/spallation signatures from the interaction  of the ultra-relativistic  QN neutrons with the preceding SN shells and surroundings (Ouyed et al. 2011).
  Applications of dsQNe to the re-ionization of the universe has been investigated in Ouyed et al. (2009a)
  while other implications to Cosmology have been put forward in Ouyed\&Staff (2011).

  The time delay between the SN and the QN, $t_{\rm delay}$, is one of
   the key parameters of the model (Ouyed et al. 2002; Ouyed et al. 2009b; Ouyed et al. 2011; 
  Ouyed\&Leahy 2012).
  QNe occurring days to weeks following the SN (dsQNe) lead to destruction of $^{56}$Ni 
    processed during the SN explosion and to the production of sub-Ni (i.e. sub-Fe) elements down 
  to lighter elements including Carbon and Lithium. 
  For very short delays, spallation lead to a complete destruction of the original $^{56}$Ni. SNe hits by the QN eject
   with delays of many weeks are most likely to survive spallation and keep their
   original $^{56}$Ni (and thus  Iron) content. These however, will experience spallation
   mostly in the outer CO-rich layers of the SN shell which as we show here lead to the formation
   of a Lithium plateau.    As it turns out, Spallation
        in the outer CO-rich  layers makes  primordial dsQNe  an efficient pre-Galactic source of 
         light elements (Be, B and Li) and of Nitrogen.
 
  Here we argue that MP stars might have 
formed from the fragmentation of  clouds polluted by the first generation of dsQNe. 
  The $^{56}$Ni depletion will leave an imprint in the material swept by the combined SN$+$QN ejecta
    which should be transmitted to the next generations of star born
     out of the fragments of the swept up material.    
  We define the shell of mass swept up  (in
  the  primordial cloud  of the progenitor star) by the mixed SN-QN ejecta as 
  $M_{\rm sw}$. At    redshift $10 < z < 20$ we adopt as our fiducial
  value  $M_{\rm sw} \sim  10^5M_{\odot}$ (Shigeyama\& Tsujimoto 1998). These swept-up clouds, as we have said,
 we expect to fragment to form the next generation of low-mass stars (e.g.  Machida et al. 2005).
  
We emphasize that in this paper we  focus on the
general idea of primordial dsQNe as plausible sources of VMP stars and of the Li plateau; only
qualitative results are presented. A more detailed investigation
will have to await  detailed simulations of the QN explosion and the
 dynamics of the collision between the the QN ejecta and the preceding/expanding SN envelope, and
  the subsequent collapse of the swept up pristine material.
 The paper is organized as follows: In \S 2, we investigate
  spallation in the inner (Ni-rich; hereafter Ni refers to $^{56}$Ni) layers of the SN while
   spallation in the outer (CO-rich) layers is described in \S 3.
   In \S 4 we discuss the formation of neutron-capture elements
    in our model and make the connection to CEMPs and NEMPs.
   The formation of the Lithium plateau is presented in \S 5
    while a general discussion and some predictions are given in \S 6.
    We conclude in \S 7.

\section{Inner-shells spallation}
\label{sec:vsn}

Here, we assume that the SN (preceding the QN)   explodes with typical kinetic energy $E_{\rm SN}$.
Following Ouyed et al. (2011), 
we  assume an onion-like profile of the resulting/expanding shocked SN ejecta (i.e., no mixing) with the innermost ejecta, viz., Ni nuclei (mass number $A_{\rm T}$=$56$)\footnote{The nickel layer/target consists of Nickel and Cobalt from Nickel decay at an amount
$N_{\rm Co,SN}= N_{\rm Ni,SN}(1-\exp{^{-t_{\rm delay}/8.8\ {\rm days}}})$. However, both
Ni and Co atoms experience spallation reducing altogether the final (i.e. total) amount of Fe produced in the cloud
swept up by the explosion.} constituting the target at a distance $R_{\rm  in}(t_{\rm delay})$=$ v_{\rm sn}\,t_{\rm delay}$ from the center of the explosion;  we assume $v_{\rm sn}$ to be constant since the QN occurs while the SN ejecta is still in the Sedov phase (e.g., McCray 1985).    In the simplest picture, 
above the Nickel layer is a Silicon layer 
and above the Si layer are  the C+O layers.
For SN progenitors in the $20M_{\odot} < M_{\rm prog.} < 40M_{\odot}$ range,  
  the amount of Nickel processed in these explosions is
   $0.1 M_{\odot} < M_{\rm Ni, SN} \le 1 M_{\odot}$  (Heger\&Woosley 2002); 
    the subscript ``SN" refers to abundances in the expanding SN shell before it gets hit by the QN ejecta. 
    Other characteristic abundance patterns of nucleosynthesis in the metal-free (Pop III) stars for the
 mass-range of interest  ($20M_{\odot} < M_{\rm prog} < 40M_{\odot}$) are given
 in Umeda et al. (2000).  For the O- and C-content in the SN shell, we take   $M_{\rm O, SN}= M_{\rm C, SN} = 1M_{\odot}$ as our fiducial values.
  We take Ni as representative of the inner shells (i.e. we ignore Si for now) and O representing the outer shells (with the C-layer overlaying the O-layer).
 We will adopt the subscript ``Ni" to  refer to the inner  layers and ``O" to
 refer to the outer  layers. We start with spallation in the inner layers.

The target number density in the Ni layer is approximately constant at $n_{\rm 56}$=$M_{\rm Ni,SN}/(4\pi R_{\rm in}^2 \Delta R_{\rm in})$, where $\Delta R_{\rm in}$ is the thickness of the Ni layer; this can be generalized to any target $A_{\rm T}$ with $n_{\rm A_{\rm T}}$=$M_{\rm A_{\rm T},SN}/(4\pi R_{\rm in}^2 \Delta R_{\rm in})$,   where $M_{\rm A_T, SN}$ is the initial mass of target $A_{\rm T}$. The neutron mean free path for spallation in the Ni layer is $\lambda_{\rm sp.}=1/(n_{\rm A_{\rm T}}\sigma_{\rm sp})$, where the spallation cross-section for neutrons on a target nucleus is empirically described with $\sigma_{\rm sp.}\simeq 45 A_{\rm T}^{0.7}$ mb (see Ouyed et al. 2011) where mb stands for  milli-barn.

The average number of collisions (i.e. the number of spallation mean-free-path $\lambda_{\rm sp.}$) an incoming neutron makes in the Ni layer is 
\begin{equation}
 N_{\rm \lambda_{\rm sp.}}\approx\frac{\Delta R_{\rm in}}{\lambda_{\rm sp.}}\simeq 3.45 \frac{M_{\rm Ni,SN}/0.5M_{\odot}}{{\left(A_{T, 56}\right)}^{0.3}(v_{\rm sn, 5000}\, t_{\rm delay, 10})^2} \ ,
\label{eq:Nsp}
\end{equation}
where $A_{\rm T,56}$ is the target atomic mass in units of $56$ and the time delay between the SN and QN explosion in units of
10 days. The velocity of the inner SN shell is in units of 5000 km s$^{-1}$.  
The expansion velocity is assumed to be constant since the SN is still
in the free expansion phase (e.g., McCray 1985) when it gets hit by the neutron-rich QN ejecta;   $t_{\rm delay}$ varies from a few days to a few weeks (see  \S \ref{sec:vsndiscuss} for other scenarios).

The  SN ejecta must be sufficiently dense that $N_{\lambda_{\rm sp.}} \geq 1$ (the condition for spallation to occur), which  limits 
the time delay between the SN explosion and the QN explosion to 
\begin{equation}
 \label{eq:tNi}
t_{\rm delay} < t_{\rm Ni} \sim 18.6\ {\rm days}\ \frac{(M_{\rm Ni,SN}/0.5M_{\odot})^{1/2}}{A_{T, 56}^{0.15}v_{\rm sn, 5000}}\ .
\end{equation}
 For longer delays ($t_{\rm delay} > t_{\rm Ni}$), spallation is minimal and Nickel and Silicon will  not be destroyed;
  note the weak dependence on the target atomic mass $A_{\rm T}$.

The average total (neutrons$+$protons) multiplicity is 
\begin{equation}
 \label{eq:zetaprimary} 
 \bar{\zeta} (E,A_{\rm T})   \simeq  7 A_{\rm T, 56} \times (1+ 0.38\ln{E}) \ , 
\end{equation}
 where the  neutron energy $E$ is in GeV.   The spallation condition, $\bar{\zeta} (E,A_{\rm T}) > 1$, yields
   a constraint on the minimum energy of the impacting nucleon
   $E > E_{\rm sp}^{\rm Ni} = 0.105$ GeV for the Nickel target.
 
 For a given energy $E$ of the incident neutrons (with $E_0$
 corresponding to the energy of the QN, first generation,  neutrons), the resulting spallation
 product's atomic weight would peak at $A_{\rm P}= A_{\rm T} -  \bar{\zeta}\sim A_{\rm T,56}\times (49 - 2.66\ln{E})$.
 A typical $E_0\sim 10$ GeV nucleon in the QN ejecta impacting
 a Nickel atom  spallates on average $\bar{\zeta}(10,56) \sim 13.12$  neutrons$+$protons.
    This  led us to posit dsQNe as the source of  Titanium  ($A_{\rm P}\simeq 56-13.12 \sim 43$) in massive stars in general
  and in Cas A in particular (see Ouyed et al. 2011)\footnote{Plausible signatures of
   the QN in Cas A have recently been suggested (Hwang\&Laming 2011).}. 
   
    The second generation of spallated neutrons+protons would have an energy $E_1= E_0/\bar{\zeta}(E_0,56)\simeq 10.0/13.2\sim 0.76$ GeV
 and so on for the subsequent generations. 
   The total, or net number of, nucleons (neutrons$+$protons) spallated per nucleon is then 
   \begin{eqnarray}
   \label{eq:multi}
   \zeta_{\rm net}&=&\prod_{i=0}^{k_{\rm max.-1}} \bar{\zeta}(E_i, A) \\\nonumber
   &=& (7 A_{\rm T,56})^{k_{\rm max.}} \times \prod_{i=0}^{k_{\rm max.-1}} (1+0.38\ln E_{\rm i})\ ,
   \end{eqnarray}
   where  $k_{\rm max} = \min(N_{\lambda_{\rm sp}}, K)$. I.e. spallation would cease when neutrons run out of target
  material ($N_{\lambda_{\rm sp.}}<1$) or when their energy is reduced below $E_{\rm sp}^{\rm Ni}$ (after $K$ spallation sublayers). 
  Details of the numerical procedure, and related references,  is described in Ouyed et al. (2011).

\begin{figure}[t!]
\begin{center}
\includegraphics[width=0.5\textwidth,height=0.5\textwidth]{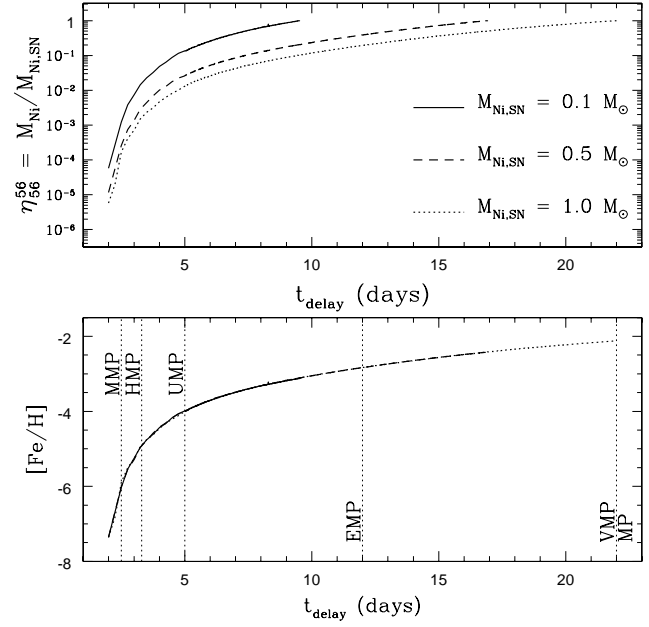}
\caption{{\bf Top panel}:   Normalized mass yield ($\eta_{56}^{56}=M_{\rm Ni}/M_{\rm Ni,SN}$; i.e. $^{56}$Ni depletion) versus time delay from spallation on the original SN $^{56}$Ni layer/target.
 {\bf Bottom panel}:  The corresponding [Fe/H]  versus $t_{\rm delay}$ (see eq. \ref{eq:FeoverH}).
  The three curves are  for $M_{\rm Ni,SN}=0.1 M_{\odot}$ (solid line),
$M_{\rm Ni,SN}=0.5 M_{\odot}$ (dashed line) and $M_{\rm Ni,SN}=1.0 M_{\odot}$ (dotted line) target mass.
 The vertical dotted lines in the bottom panel demarcate the different classes of metallicity; MP, VMP,
EMP, UMP, HMP and, MMP stars as defined in Beers\&Christlieb (2005).}
\label{fig:FeH}
\end{center}
\end{figure}

\subsection{[Fe/H] trends}

%  Solar abundances
% $log{\epsilon_{\rm Fe}} = 7.5 $
% $log{\epsilon_{\rm Si}} = 7.51 $
% $log{\epsilon_{\rm O}} = 8.69 $
% $log{\epsilon_{\rm N}} = 7.83 $
% $log{\epsilon_{\rm C}} = 8.43 $
% $log{\epsilon_{\rm Li}} = 1.05 $

We define the  normalized mass yields of spallated fragments  of atomic
 mass $A$ as $\eta_{\rm A}^{\rm A_T}= M_{\rm A}/ M_{\rm A_T, SN}$.
  The superscript refer to the target while the subscript refers to the spallation product.
   For example, $\eta_7^{56}$ would correspond to $^7$Li
   produced from spallation on a Nickel target. In this definition,   
   $\eta_{56}^{56}$ would correspond to reduction of the initial (i.e. SN)
   Nickel content.

Figure \ref{fig:FeH} shows Nickel destruction, i.e.   $\eta_{56}^{56}$, versus $t_{\rm delay}$ for three different
 initial Nickel content, $M_{\rm Ni, SN}=0.1,0.5$ and 1$M_{\odot}$. 
      The corresponding  [Fe/H] abundance  (shown in the bottom panel of Figure 1) is derived from
 \begin{eqnarray}
  \label{eq:FeoverH}
 \left[\frac{Fe}{H}\right]  
&=&  \log{\frac{N_{\rm Fe}}{N_{\rm Fe, SN}}} +   \log{\frac{N_{\rm Fe, SN}}{N_{\rm H}}} - \log{\frac{N_{\rm Fe}}{N_{\rm H}}}\vert_{\odot} \\\nonumber 
&\simeq&   \log{\eta_{\rm 56}^{56}} +   \log{\frac{(M_{\rm 56, SN}/ 0.5  M_{\odot})}{(M_{\rm sw}/10^5M_{\odot})}} -2.42\ ,
\end{eqnarray}
 where  the number of atoms of elements A is  $N_{\rm A} = M_{\rm A}/A$. 
Here $M_{\rm sw}$ is the mass swept up by the  mixed SN-QN ejecta  in
  the  primordial cloud (at redshift $10 < z < 20$; near the end of the dying out of the first stars) of the progenitor star which we take to be   
of the order of  $M_{\rm sw} \sim  10^5M_{\odot}$ (Shigeyama\&Tsujimoto 1998). We adopt 1.2 as the mean molecular weight of the neutral primordial gas. The evolution and cooling  of the swept-up cloud
 we expect to fragment to form the next generation of stars (e.g.  Machida et al. 2005). 
    
  If the swept material is mixed
 well with the ejecta  of the dsQN (i.e. complete mixing),
  the metal abundance in the gas shell (and thus the next stars)   should carry
 a ``genetic imprint" of the ``iron impoverishment" and spallation products from the dsQN explosion.

As can be seen in the bottom panel of Figure \ref{fig:FeH}, 
 the range in metallicity we obtain is  $-7.5 < {\rm [Fe/H]} < -2.0$ for $ 2\le t_{\rm delay} ({\rm days}) < 25$.
  For a given  $M_{Ni, SN}$, QN occurring with $t_{\rm delay} << t_{\rm Ni}$ 
 lead  to important depletion of Nickel.  In our model, SN progenitors
 with the lowest initial Ni content experiencing the QN with short delays
 would lead to  MMP stars in the nomenclature given in Beers\&Christlieb (2005).
   Progenitors with highest initial Ni content hit by the QN ejecta
   with $t_{\rm delay} > t_{\rm Ni}$ will experience minimal spallation
   and would correspond to MP stars.  
QNe progenitors with even higher Ni content (i.e.
processed during the SN proper; e.g.  $M_{\rm Ni, SN}> 1 M_{\odot}$) but who have not
experienced excessive spallation (i.e. $t_{\rm delay}> t_{\rm Ni}$  or equivalently $\eta_{\rm 56}^{56}\sim 1$) would
lead to  [Fe/H]$>-2$.  

The range of metallicity we arrive at  resemble those of the  MP stars in the
   Galactic halo (e.g.  Beers et al. 1992) and also accounts for HMP stars 
  such as HE 0107$-$5240 ([Fe/H]$\sim -5.3$; Christlieb et al. 2002)
  and HE 1327$-$2326 ([Fe/H]$\sim -5.4$; Frebel et al. 2005).
The lower limit,  [Fe/H]$\sim -7.5$, corresponds to  QNe progenitors with low
iron content in the SN proper ($\sim 0.1M_{\odot}$) but have experienced
excessive spallation (i.e. $t_{\rm delay}<<  t_{\rm Ni}$ days with a corresponding $\eta_{\rm 56}^{56}\sim 10^{-5}$).
For $t_{\rm delay} < 2$ days the metallicity drops to even much lower values as can be derived
 from extrapolation to smaller $t_{\rm delay}$ in the bottom panel of Figure 1.

  Machida et al. (2005) point out that  HMP stars has too little metallicity to be explained by
   the mixture of SN ejecta of a first-generation massive star and the primordial gas swept
   in its SNR shell.  For SN models to explain HMP stars metallicity, the ejecta gas needs to sweep  up much more
   ($> 100 M_{\rm sw}$) primordial gas which seems very unlikely. In our model,
    these correspond to primordial dsQNe with $t_{\rm delay} < 4$ days.
   Finally,  we note that the above analysis, assume complete mixing between the swept ambient gas and the dsQN ejecta. 
   The individual stars, however, expected to show a large scatter since the efficiency of mixing may vary from fragments to fragments.

  \begin{figure*}[t!]
\begin{center}
\begin{tabular}{cc}
\includegraphics[scale=0.72]{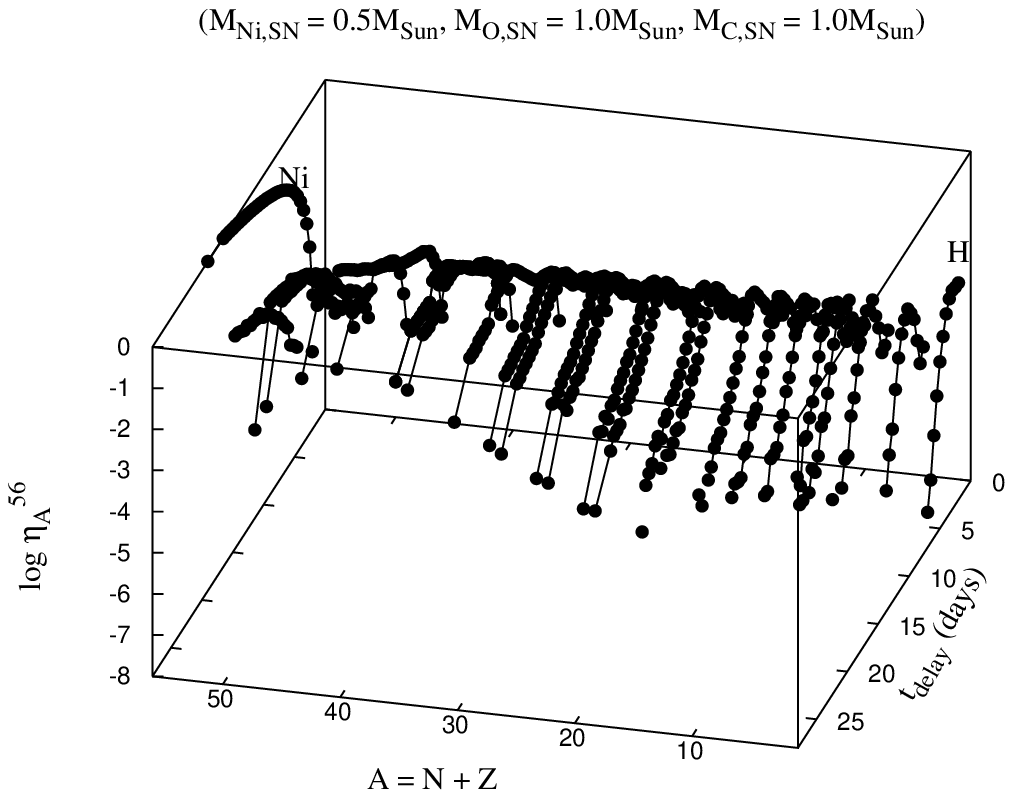} & \includegraphics[scale=0.72]{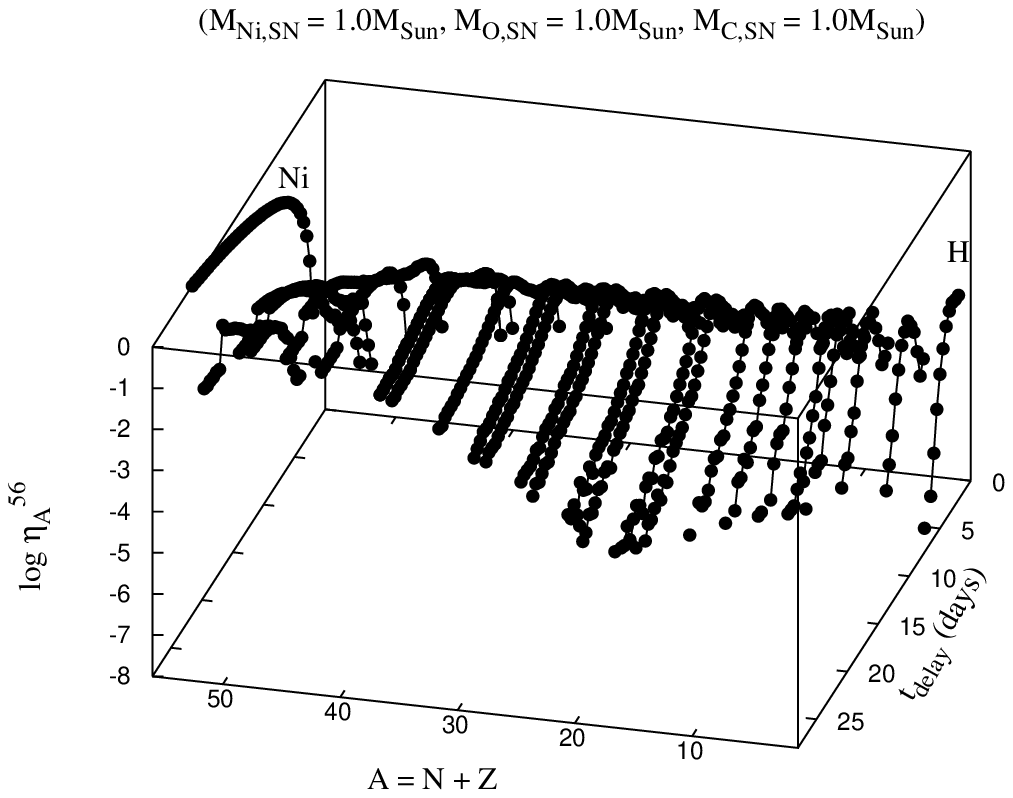}\\
\includegraphics[scale=0.72]{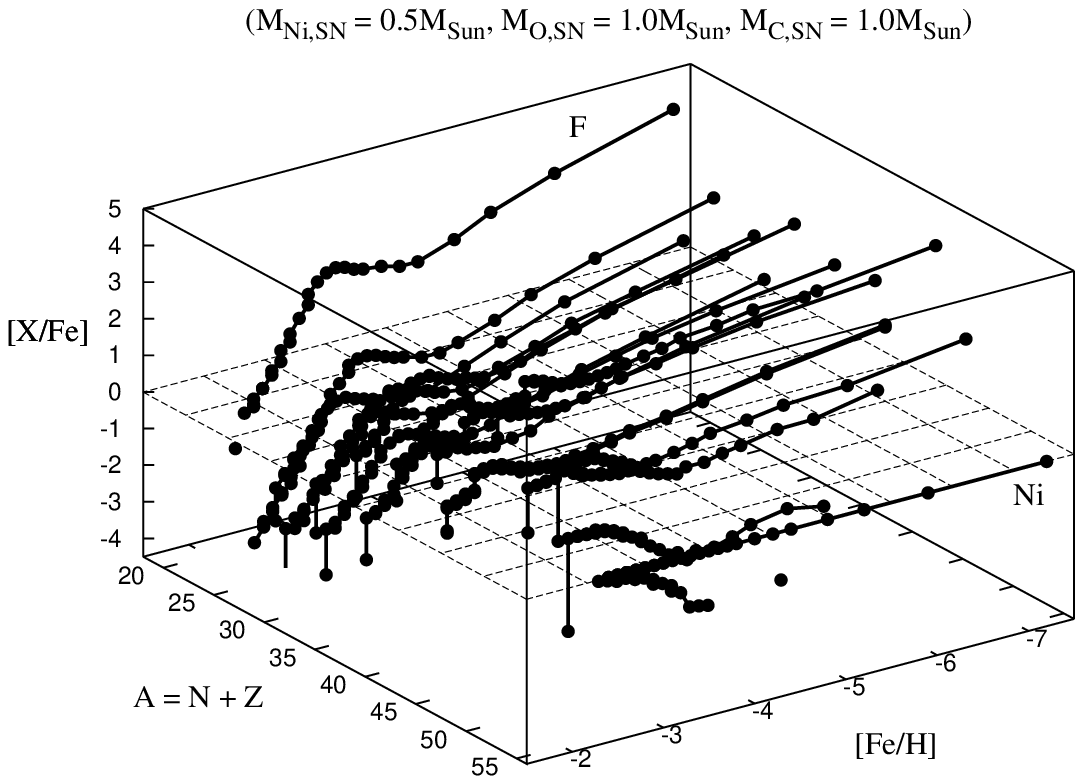} & \includegraphics[scale=0.72]{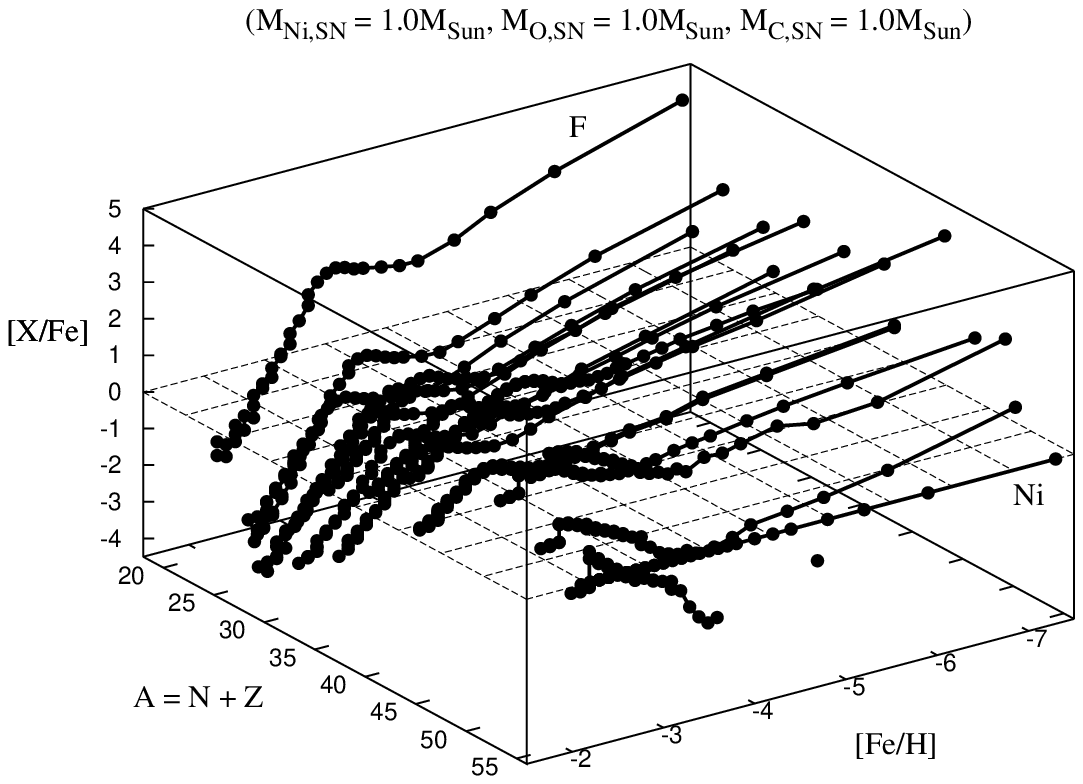}\\
\includegraphics[scale=0.72]{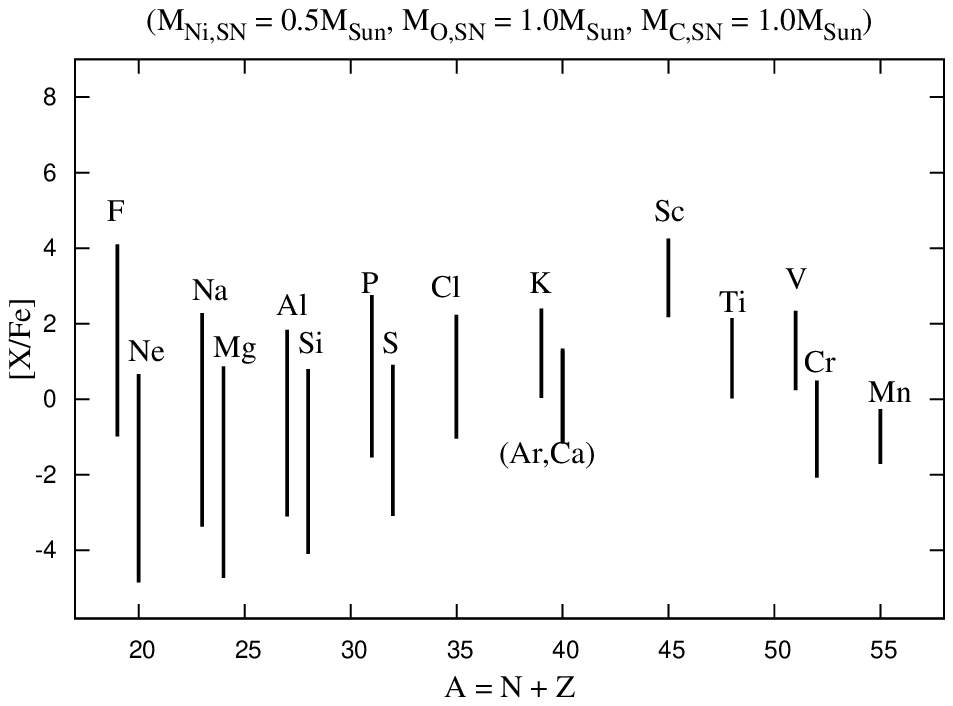}  & \includegraphics[scale=0.72]{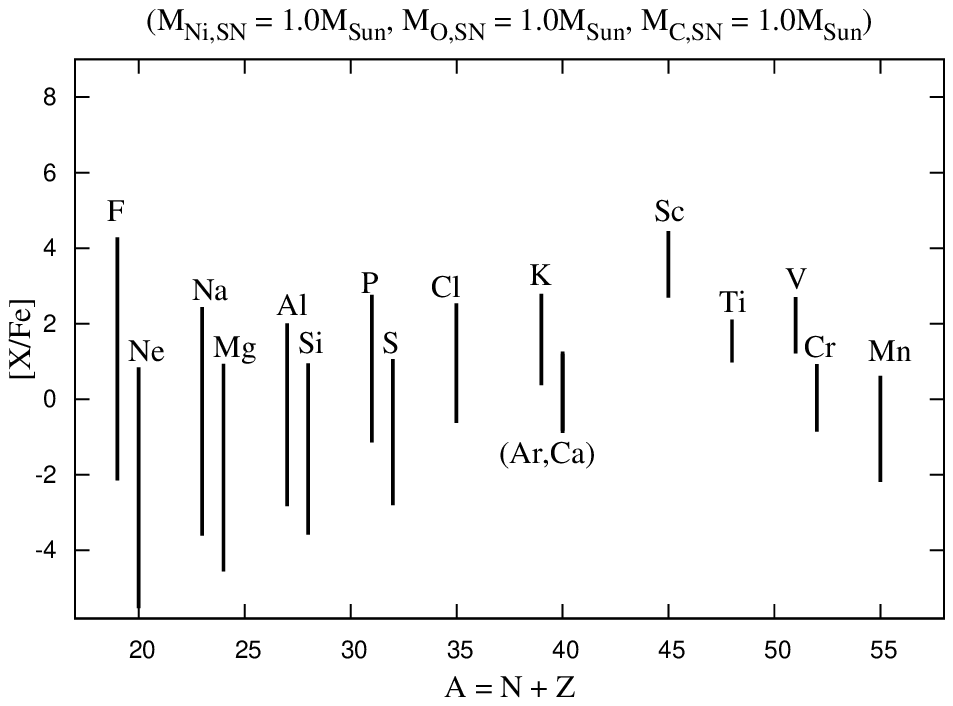}
\end{tabular}
\caption{{\bf Top panels}:  Mass yields of Ni-spallation products ($\log{\eta_{\rm A}^{56}}= \log{M_{\rm A}/M_{\rm Ni,SN}}$ with atomic number $1 \le A \le 56$) versus time delay ($t_{\rm delay}$ in days). All stable isotopes down to H are shown
with the Ni  and H curves indicated  for reference. 
{\bf Middle panels}:  [X/Fe] versus the corresponding  [Fe/H] for products with atomic number  $16 < A \le 56$.
 {\bf Bottom panels}: [X/Fe]) versus
 atomic number   ($16 < A \le 56$) obtained from projecting  the curves  in the middle
 panels into the (A,[X/Fe]) plane. 
 Left panels are  for $M_{\rm Ni,SN}=0.5 M_{\odot}$ while right panels are for 
$M_{\rm Ni,SN}=1.0 M_{\odot}$.} 
\label{fig:XoverFe-in}
\end{center}
\end{figure*}

 \begin{figure*}[t!]
\begin{center}
\begin{tabular}{cc}
\includegraphics[scale=0.72]{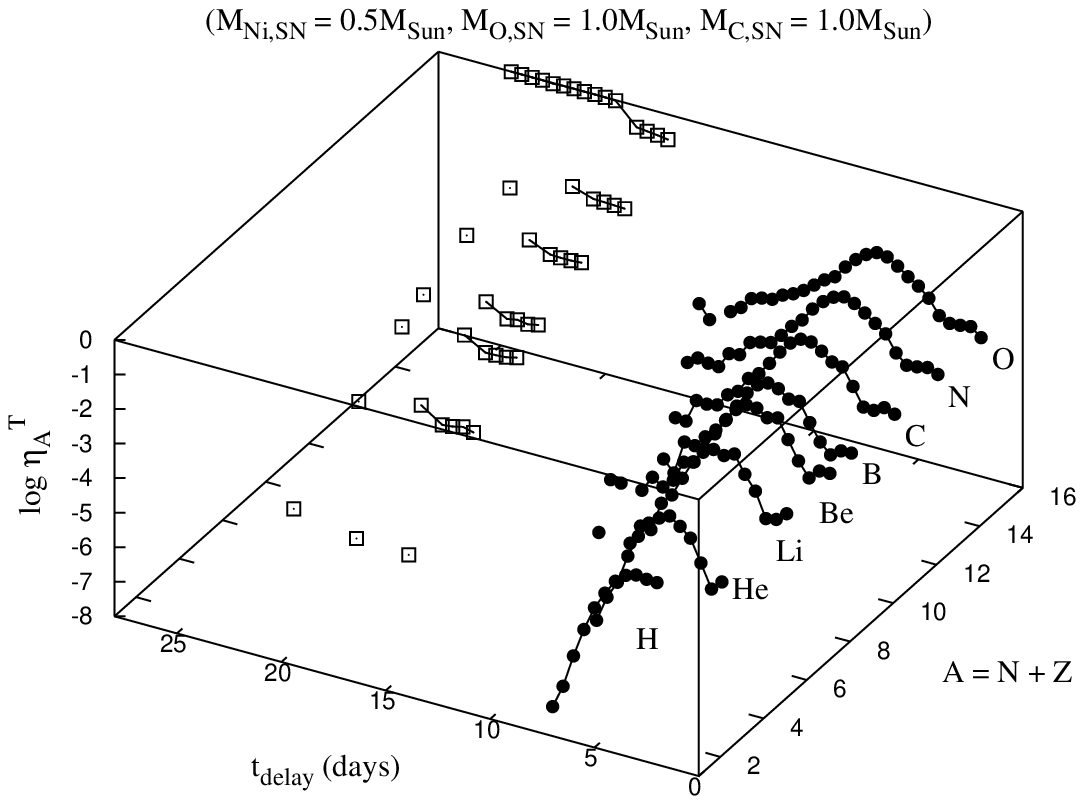} & \includegraphics[scale=0.72]{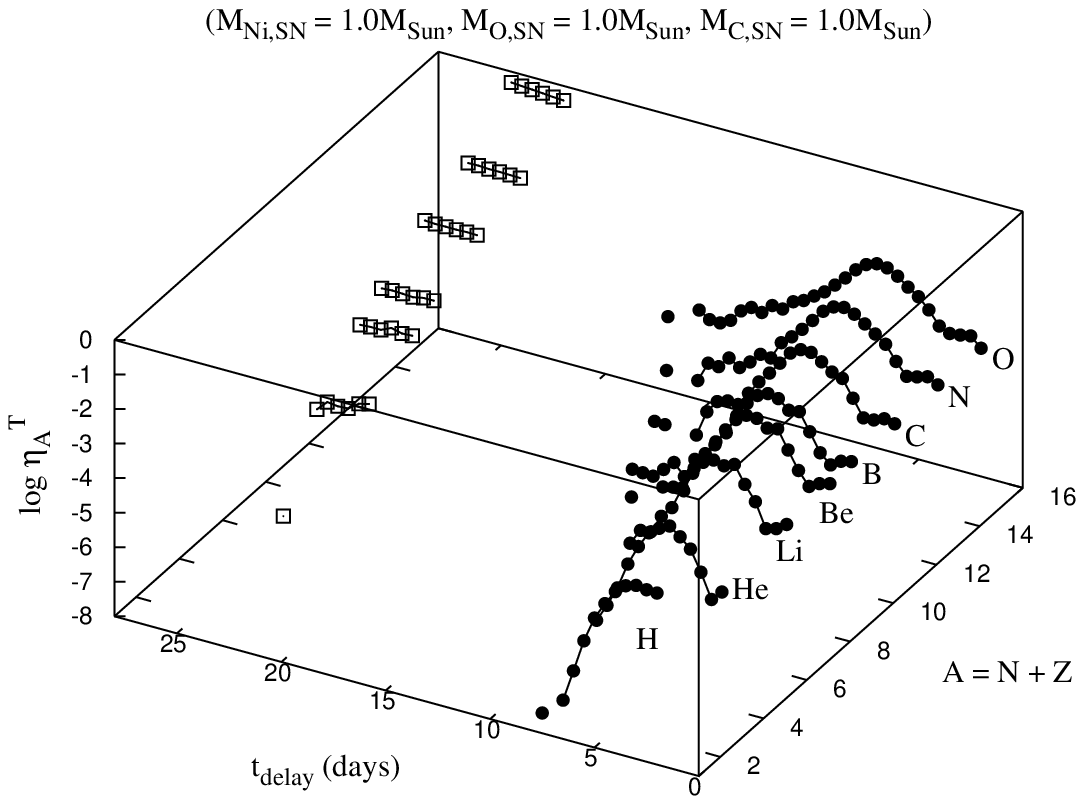}\\
\includegraphics[scale=0.72]{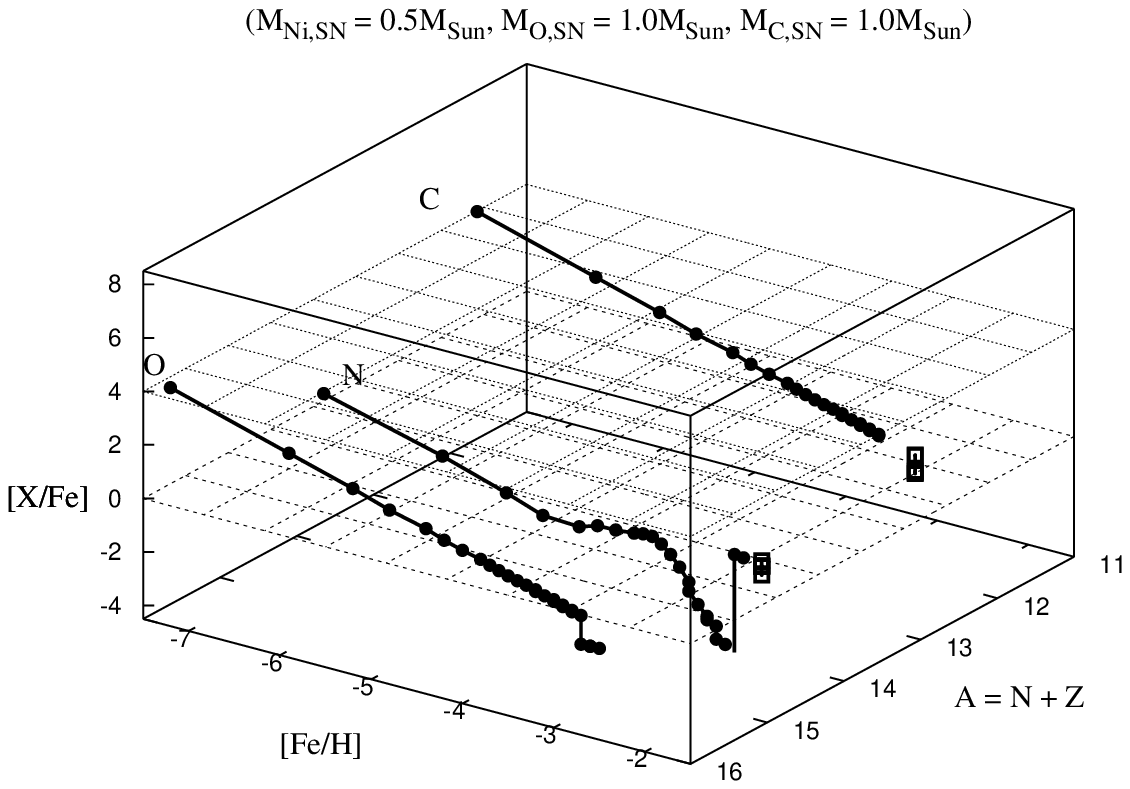} & \includegraphics[scale=0.72]{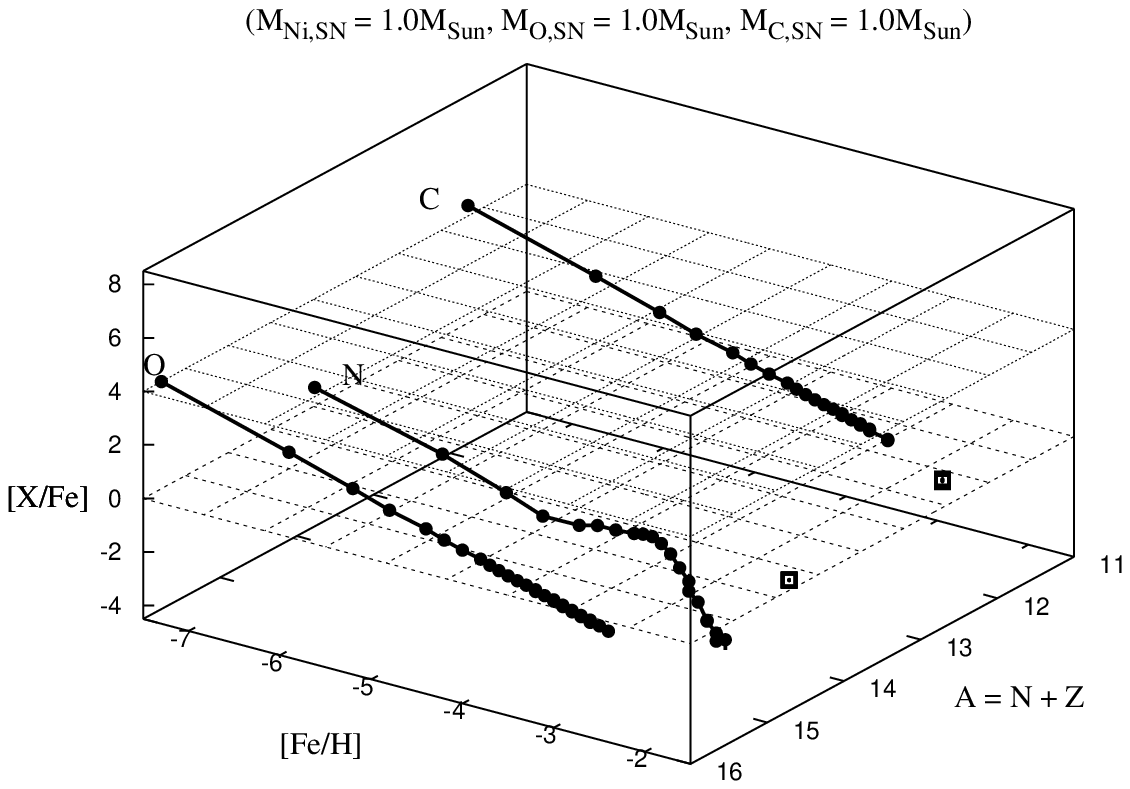}\\
\includegraphics[scale=0.72]{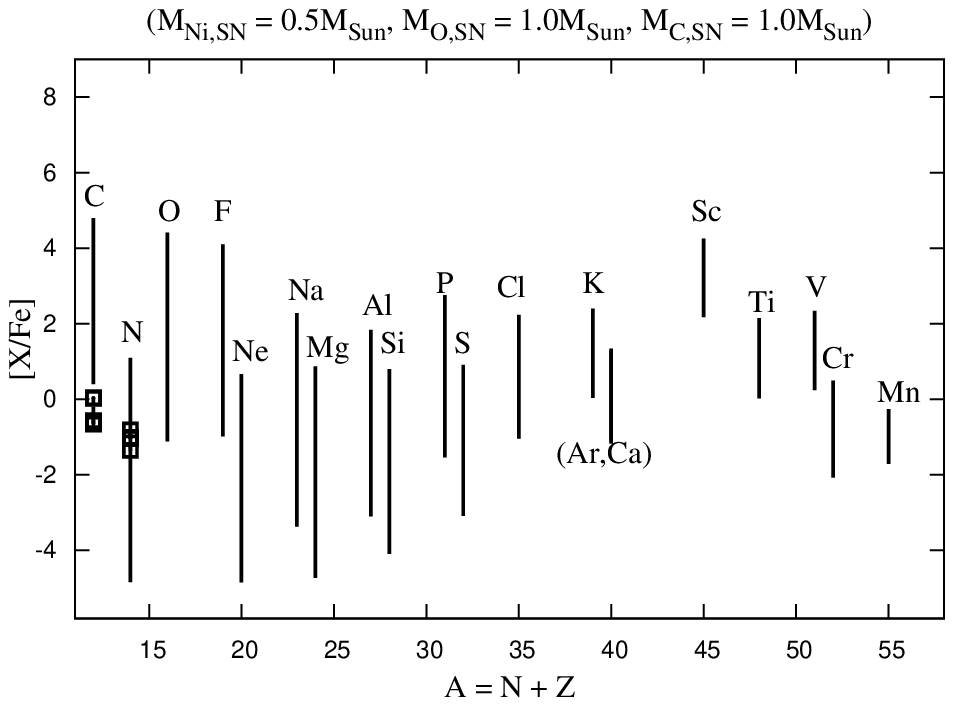}  & \includegraphics[scale=0.72]{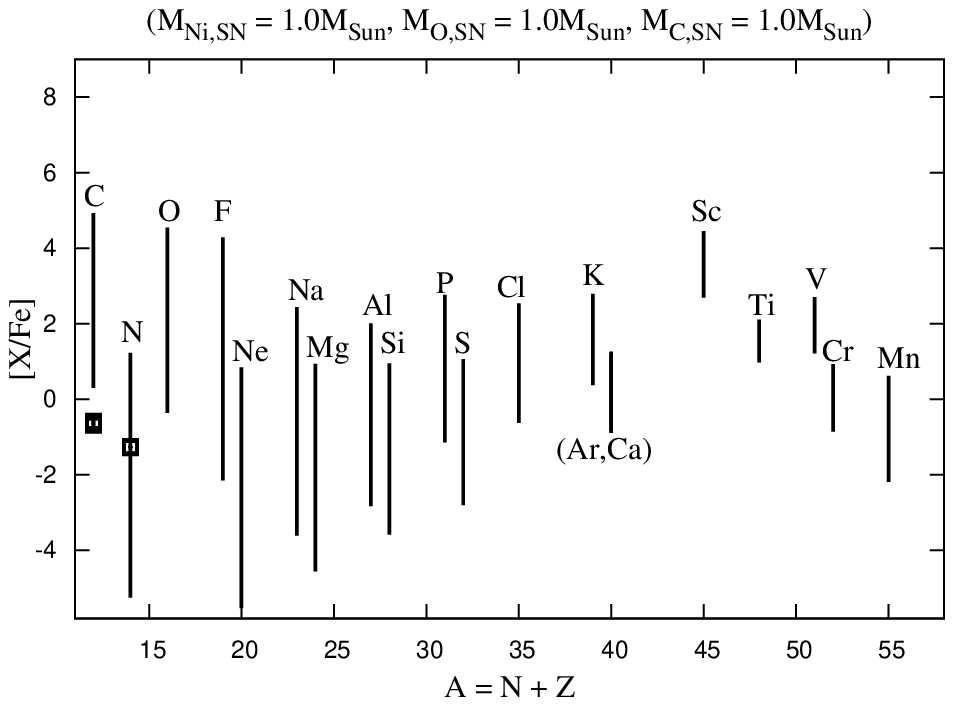}
\end{tabular}
\caption{ {\bf Top panels}:   Comparison of  Ni-spallation (solid circles) and O-spallation (open squares)
 mass yields  versus  time delay ($t_{\rm delay}$ in days) for  $1\le A \le 16$. All stable isotopes down to H are 
labelled.  {\bf Middle panels}:  [X/Fe] versus 
 the corresponding  [Fe/H] in our model.   Only O,N, and C are shown here with 
  the two horizontal planes, shown for reference, corresponding to [X/Fe]=0 and [X/Fe]=4.
 {\bf Bottom panels}: [X/Fe] versus 
 atomic number  A for all stables isotopes $12\le A \le 56$.   [C/Fe] and
 [O/Fe] from O-spallation are shown as open squares while [X/Fe] from Ni-spallation are shown as
  solid lines. Left  (right) panels are  for $M_{\rm Ni,SN}=0.5 M_{\odot}$ ($M_{\rm Ni,SN}=1.0 M_{\odot}$).
} 
\label{fig:XoverFe-out}
\end{center}
\end{figure*}

\subsection{[X/Fe] trends}

 The top panels in Figure \ref{fig:XoverFe-in}  show normalized mass-yields, $\eta_{\rm A}^{\rm A_T}$,
  for stable isotopes with atomic weight $1\le A \le 56$ and for different $t_{\rm delay}$;
  the left (right) panels are for $M_{\rm Ni,SN}=0.5M_{\odot}$ ($M_{\rm Ni,SN}=1.0M_{\odot}$). 
  As expected,  for the shortest delays spallation is substantial, converting
  most of the Ni into H.  For $ 2\ {\rm days} < t_{\rm delay} < 10\ {\rm days}$,
   almost all of the stables isotopes are produced.
   The middle  panels show the corresponding  [X/Fe]  
 versus  [Fe/H] (see Appendix for the derivation of [X/Fe] in our model).  In these middle panels, only
 $A > 16$ are shown since  elements below (and including) O will
 be altered by spallation in the outer layers (these will be discussed  in the next section).

  In primordial dsQNe, inner-shells spallation  depletes Ni  (and thus the iron content in the swept cloud) and enhances the abundances of  sub-Fe metals.
 Naturally,  among the spallation products (see Figure \ref{fig:XoverFe-in}) are   all stable isotopes of K, Ti, V, Cr, Mn, and Fe plus any unstable isotopes that decay into these.  Interestingly iron peak elements ($22 < Z < 28$, i.e. Ti, V, Cr, Mn, Fe, Co, Ni) have been  observed in MP stars.
 We note that these elements appear to be difficult to synthesize in standard
 nucleosynthesis models of SNe (Kobayashi et al. 2006; 
  e.g. Yoshida et al. 2008 for alternatives).  Thus our model provides a different origin of these elements
  in MP stars and seems to account for the overall observed trends.
For example, spectroscopic studies report   increasing [Ti/Fe] ratios with decreasing metallicity.
 This finds a natural explanation in our model since  in general Ni spallation
   results in enhanced abundances relative to solar of the sub-Fe elements such as  Ti.  We find that for
  $ 2\ {\rm days} < t_{\rm delay} < 10\ {\rm days}$, elements closer to Ni (e.g. Cr and  Mn) on average   decrease with
 decreasing metallicity while lower $A$ elements (e.g. V)  show the opposite trends.  
         The bottom panels in Figure \ref{fig:XoverFe-in}  show [X/Fe] projected into the (A,[X/Fe]) plane.
   Spallation in the inner Ni-layer produces     on average more Fluorine and Scandium than 
     the other sub-Fe elements  (see bottom panels in figures \ref{fig:XoverFe-in} and \ref{fig:XoverFe-out}).
     As discussed later in \S \ref{sec:FandSc}, these elements could be a telltale signature
     of dsQNe in Galactic halo MP stars.

We close this section with a brief discussion of our findings:

\begin{itemize}

\item In our model, the [Fe/H]  in the swept cloud depends on the
 initial Nickel mass (i.e. most likely on the mass of the SN progenitor) and the time delay between
  the SN and the QN.  There are two classes/regimes  of dsQNe that could give similar mass
  yields in [Fe/H]. 
    dsQNe impacting SN shells from massive  progenitors (presumably those with high initial Nickel content)
    with short delays   and  dsQNe impacting SN shells from less massive  progenitors 
    with long delays.    However, these two cases  differ in their final abundances of the sub-Ni products.
     This means that the  low-mass  stars born from the  clouds swept-up by these two classes of dsQNe, although
        similar in iron abundance,   might exhibit 
     a variety of different chemical abundance patterns.
      It would be misleading to conclude that  Galactic halo MP stars with similar [Fe/H] values but
      showing difference in chemical abundances, do not  share a common origin.
 
 \item    While  one is tempted to associate low-Fe abundance with earlier times,
      our model provides a fundamentally different view which separates
       the iron content in low-mass Galactic Halo stars from evolution and thus age. dsQNe
       provide a common origin and time for the formation of MP stars
        without the need for multiple events. I.e.  high metallicity in halo stars does not imply
        evolution and/or multiple events according to our model. 
        
\end{itemize}

\section{Outer-shells spallation}
\label{sec:outer}

We now take into account the presence of O and C (but no N) in the outer layers
 of the SN shell prior to the QN impact.   For short time delays (with important Ni spallation), spallation in the outer layers is minimal
   and the original (SN)  C and O content is preserved. 
   As we show below, this effectively corresponds to a very pronounced  O- and C-enhancement 
    at shortest delays (i.e. lowest  [Fe/H] values in our model). In this section, 
    we explore the conditions under
 which spallation proceeds into these layers and the consequences
 on abundances of $A\le 16$ elements.  We first focus on C, O and N while Li 
  is discussed in \S 4. Ba and  Be  are briefly discussed in \S 5.

\subsection{Trends in Oxygen}

The spallation condition, $\bar{\zeta}(E,16) > 1$, yields
   a constraint on the minimum energy of the impacting nucleon
   $E> E_{\rm sp}^{\rm O} = 0.268$ GeV for an Oxygen target (i.e. $A_{\rm T}=16$). Thus 
 for $t_{\rm delay} < t_{\rm Ni}$, 
 a necessary condition for spallation to continue onto the  outer layers  is 
 $E_0/\zeta_{\rm net} >  0.268$ GeV with $\zeta_{\rm net}$  given in eq(\ref{eq:multi}).
A reasonable approximation  is $\prod_{i=0}^{k_{\rm max.-1}} (1+0.38\ln E_{\rm i})\sim (1+0.38\ln E_{\rm av.})^{k_{\rm max.}}$
   where $\ln E_{\rm av.}= (\ln E_{\rm 0} + \ln E_{\rm sp}^{\rm O})/2\sim 0.492$ for our fiducial value of $E_0=10$ GeV.
   This gives $\zeta_{\rm net}\sim (7 A_{\rm T,56})^{k_{\rm max.}}\times (1.19)^{k_{\rm max.}}$ which means
    that the $E_{\rm sp}^{\rm O}=  E_0/\zeta_{\rm net} > 0.268$ GeV condition translates to 
  a condition on the number of spallation layers
  in the inner SN shells to be $k_{\rm max} < \ln{(37.32 E_{\rm 0, 10 GeV})}/\ln{(8.33 A_{T, 56})}$.
  For $N_{\lambda_{\rm sp}}\sim k_{\rm max}$,  and using Eq.(\ref{eq:Nsp}), we arrive at
   \begin{equation}
    \label{eq:tO}
   t_{\rm delay} > t_{\rm O}\sim 14\ {\rm days}\ \frac{(M_{\rm Ni, SN}/ 0.5M_{\odot})^{1/2}}{A_{T,56}^{0.15} v_{\rm sn, 5000}}\ .
   \end{equation}
   This is the second critical timescale in our model (also valid for C) which defines a regime
    where spallation occurs both in the inner (Ni) and outer (O) shells. The expression
    above  agrees well with the numerical values (see Table 1).

    For $t_{\rm delay} > t_{\rm O}$ we isolate two regimes of 
    spallation on the O target:   
   \begin{equation}
\bar{\zeta}\sim 
 \left\{ 
 \begin{array}{rl}
 1.8 &\mbox{if $ t_{\rm O} < t_{\rm delay}  < t_{\rm Ni}$\ since\ $E_1 \sim 0.76\ {\rm GeV}$,} \\
  3.8 &\mbox{if $ t_{\rm delay}  > t_{\rm Ni}$\ since\ $E_1\sim E_0= 10\ {\rm GeV}$,}
       \end{array}
 \right.
 \end{equation} 
which means that O-spallation would lead mainly to Carbon ($A_{\rm P}= 16-3.8\sim 12$) when $t_{\rm delay}> t_{\rm Ni}$
(i.e. when spallation occurs mainly in the outer SN shell) and Nitrogen
($A_{\rm P}= 16-1.8\sim 14$) formation when spallation starts in the inner shells and proceeds
to the outer ones; i.e. when $  t_{\rm O} <t_{\rm delay} < t_{\rm Ni}$.

   The top panels in Figure \ref{fig:XoverFe-out} show normalized mass-yields, $\eta_{\rm A}^{\rm A_T}$,
  for stable isotopes ($1\le A \le 16$) and for different $t_{\rm delay}$.  The mass-yields
  from the inner-shell spallation ($\eta_{\rm A}^{\rm 56}$) are shown as solid circles while the mass-yields
  from the outer-shell spallation ($\eta_{\rm A}^{\rm 16}$) are shown as open squares. 
  As can be seen from equation (\ref{eq:tO}) above, the more Ni present in the SN shell, 
    the longer $t_{\rm O} $  before spallation proceeds
   onto the outer layers; $t_{\rm delay} > t_{\rm O} \sim 14$ days
   for  $M_{\rm Ni,SN}=0.5M_{\odot}$ and  $t_{\rm delay} >t_{\rm O} \sim 19.8$ days
   for $M_{\rm Ni,SN}=1M_{\odot}$. This  explains the gaps between inner-spallation
    and outer-shell spallation regimes in the top panels of figure  \ref{fig:XoverFe-out} .

   The middle   panels show the corresponding [X/Fe]
 versus  [Fe/H] for O, N and C.  The two horizontal planes
 correspond to [X/Fe]=0 and [X/Fe]=4.  There is a clear enhancement
 in O and C abundance compared to Iron with decreasing metallicity (i.e. decreasing $t_{\rm delay}$)
  in our model.   For $t_{\rm delay} < t_{\rm O}$, spallation does not occur
    in the outer (CO) layers while Nickel (and thus later Fe in the swept up cloud) is destroyed.
    This lead to the high values of [O/Fe] and [C/Fe] as explained
    further in \S \ref{sec:carbon}.   Nitrogen is less enhanced
  than O and C since N has to be first produced by spallation 
  in the O layers (see \S \ref{sec:nitrogen}).   
    The bottom panels show the (A,[X/Fe])  plane with 
  the [X/Fe]  contribution from the outer-shell spallation  (shown as the empty squares)
  is compared to the [X/Fe] from the Ni-spallation (solid lines).

    \subsubsection{The plateaus in sub-O spallation products}
    \label{sec:plateaus}
    
   The plateaus in N, C, B, Be, Li and He from outer-shell spallation (top panels in  figure \ref{fig:XoverFe-out})
     can be understood as follows: 
         The $t_{\rm delay} > t_{\rm O}$ condition given by eq. (\ref{eq:tO})
     combined with eq.(\ref{eq:Nsp}) gives $N_{\lambda_{\rm sp}} < 1.8$.
     This means that in $t_{\rm delay} > t_{\rm O}$ dsQNe,  the thickness
      of the Ni layer is of the order of one neutron-spallation mean-free-path. 
         Effectively, the QN neutrons interact  on average once inside the Ni-layer. 
         The resulting average multiplicity is $\bar{\zeta}(10,56)\sim 13.2$ for our fiducial values
          with most of the  spallated neutrons+protons heading towards the O-layer with an 
          energy $E_0/13.2\sim 0.76$ GeV.  This means that $\bar{\zeta}(0.76,16)\equiv \bar{\zeta}^{\rm O}\sim 1.8$.         
            Thus spallation in the O-layer
           would lead to a product peaking at $A_{\rm P} = 16 - \bar{\zeta}^{\rm O}\sim 16-1.8\sim 14$
            for any dsQNe with  $t_{\rm O} < t_{\rm delay} < t_{\rm Ni}$.  The peak in spallation products at N
             and  the weak dependence on  $t_{\rm delay}$ (as well
             as the narrow window $t_{\rm O} < t_{\rm delay} < t_{\rm Ni}$) explains
              the ``descending step-ladder" in the  mass-yield $\eta_{\rm A}^{16}$ (and thus abundances)
             of sub-O elements in the top panels of  figure \ref{fig:XoverFe-out}.

\subsection{Trends in Carbon}
\label{sec:carbon}

\begin{figure*}[t!]
\begin{center}
\includegraphics[width=0.495\textwidth, height=0.45\textwidth]{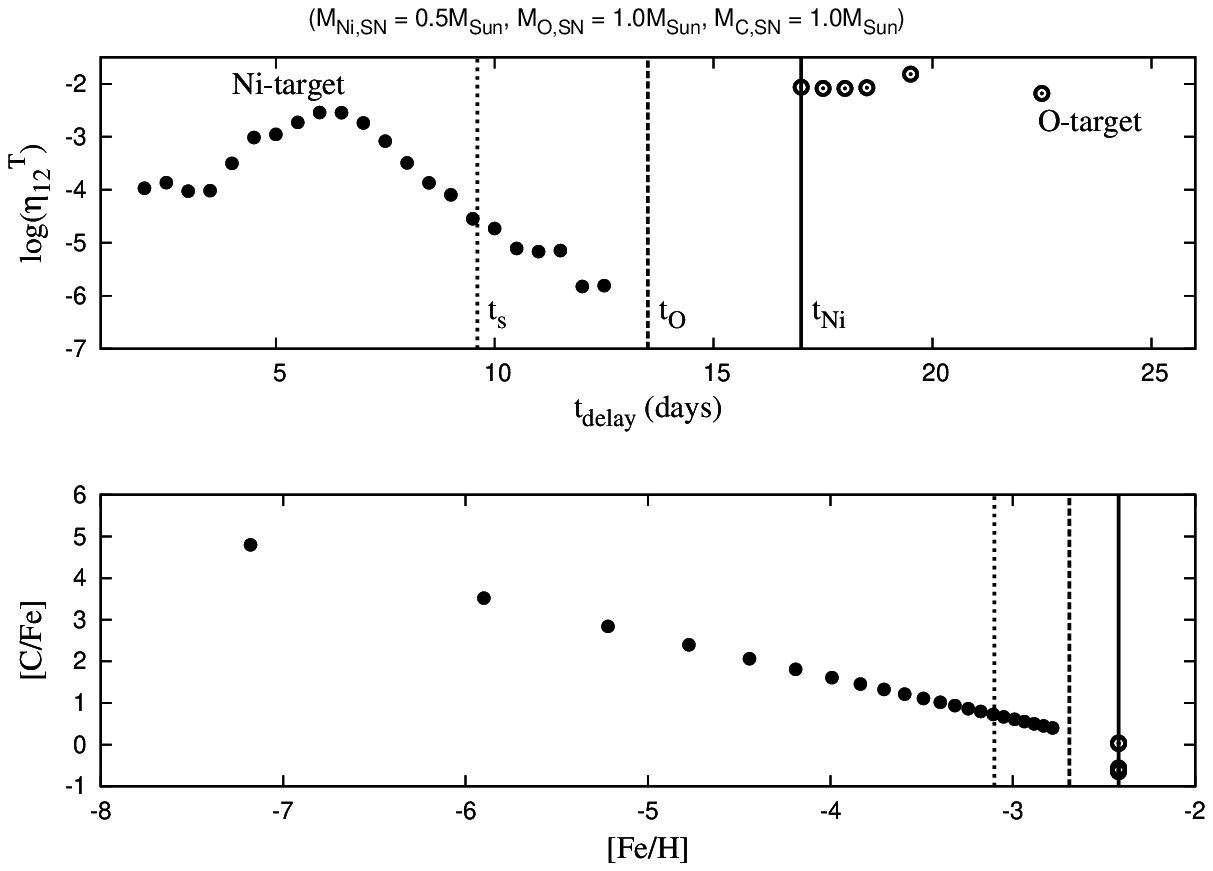}
\includegraphics[width=0.495\textwidth, height=0.45\textwidth]{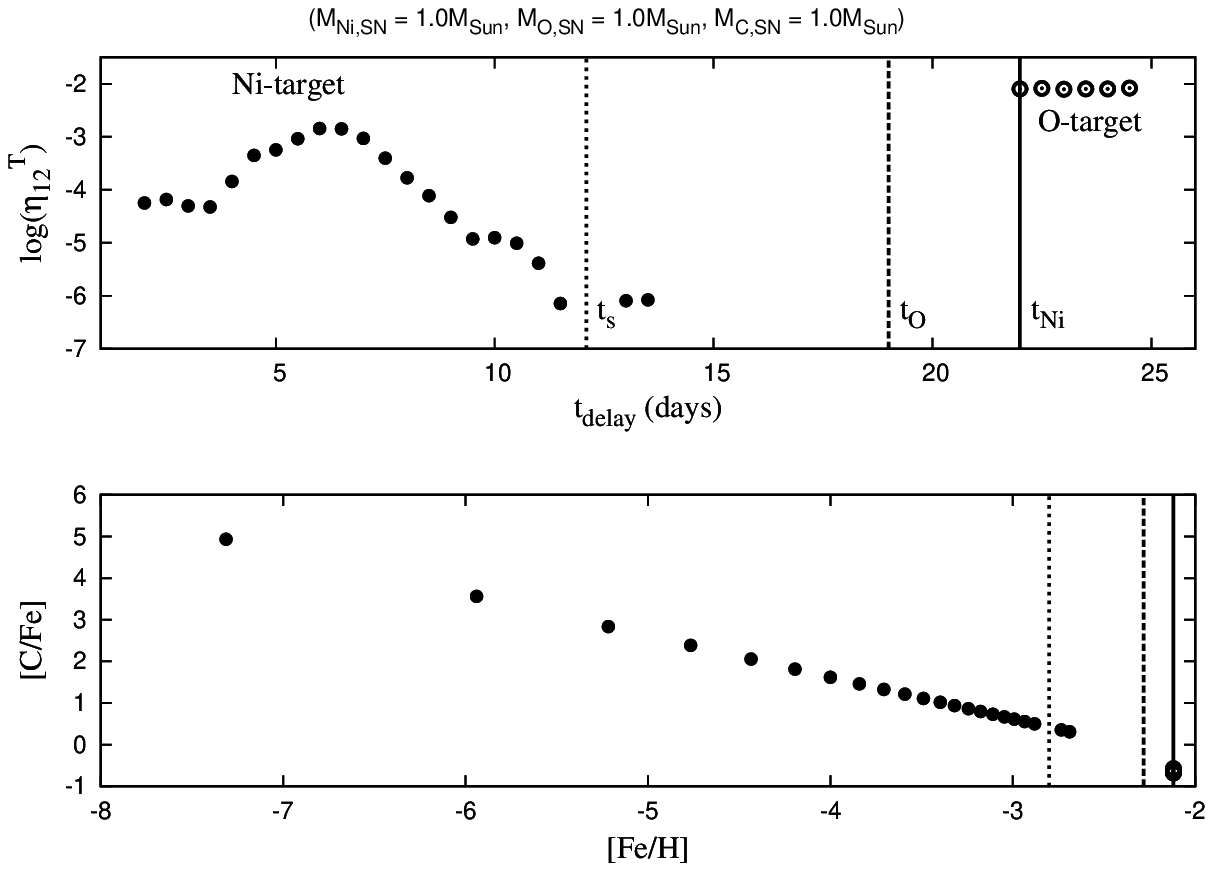}
\caption{{\bf Top panels}:  Carbon mass yields  versus
time delay ($t_{\rm delay}$) from Ni-spallation  ($\eta_{12}^{56}=M_{\rm C}/M_{\rm Ni,SN}$; solid circles)
 and  O-spallation   ($\eta_{12}^{16}= M_{\rm C}/M_{\rm O,SN}$; open circles).
{\bf Bottom Panels}: The corresponding [C/Fe] versus [Fe/H]. Left panels are  for  ($M_{\rm Ni,SN}=0.5 M_{\odot}, M_{\rm O,SN}=1.0 M_{\odot}, M_{\rm C,SN}=1.0 M_{\odot}$) while right panels are for 
($M_{\rm Ni,SN}=1.0 M_{\odot}, M_{\rm O,SN}=1.0 M_{\odot}, M_{\rm C,SN}=1.0 M_{\odot}$). 
 The vertical lines labelled $t_{\rm Ni}, t_{\rm O}$, and $t_{\rm s}$ (see Table 1)  represent demarcations between
  purely Ni- (i.e. inner shell)-spallation for $t_{\rm delay} < t_{\rm O}$ and 
  purely O- (i.e. outer shell)-spallation for $t_{\rm delay} > t_{\rm Ni}$.
  For $t_{\rm O} < t_{\rm delay} < t_{\rm Ni}$ spallation occurs in both the inner
  and outer SN layers. The dotted vertical line defines  the
     neutron-capture (s-process) regime ($t< t_{\rm s}$; see \S \ref{sec:sprocess}). For $t_{\rm delay}>t_{\rm Ni}$  Ni-spallation  is minimal 
 which  means that points to the right of the solid vertical line in the top panels
 all fall on the solid line in the bottom panels;  i.e. all have the same value of  [Fe/H] as given by eq. (\ref{eq:FeoverH})
  when $\eta_{56}^{56}\sim 1$. These vertical
     lines are shown in all  subsequent figures.}
\label{fig:CoverFe}
\end{center}
\end{figure*}

The top panels in Figure \ref{fig:CoverFe} compares C yield from Ni- and O-spallation.
The critical times $t_{\rm Ni}$ (solid vertical line) and $t_{\rm O}$ (dotted vertical lines) are also shown.
In general C produced from Ni-spallation is much less than C produced
from O-spallation.  The initial (SN) C experiences spallation into lighter elements for $t_{\rm delay} > t_{\rm O}$
 but the residual C still exceeds the Ni-spallated C.  
As derived in the appendix, we get (for $t_{\rm delay} < t_{\rm Ni}$)
\begin{eqnarray}
{\rm \left[\frac{C}{Fe}\right]_{\rm Ni}} &\simeq&  \log{\frac{\eta_{\rm 12}^{56}}{\eta_{\rm 56}^{56}}} - 0.26 \\ \nonumber 
&+& \log{\left( 1 + \frac{\eta_{\rm 12}^{16}}{\eta_{\rm 12}^{56}}\frac{M_{\rm O,SN}}{M_{\rm Ni,SN}} 
 +  \frac{\eta_{\rm 12}^{\rm 12}}{\eta_{\rm 12}^{56}}\frac{M_{\rm C,SN}}{M_{\rm Ni,SN}} \right)} 
\end{eqnarray}
and (for $t_{\rm delay} > t_{\rm Ni}$)
\begin{eqnarray}
{\rm \left[\frac{C}{Fe}\right]_{\rm O}} &\simeq& \log{\eta_{\rm 12}^{16}} - 0.26\\\nonumber 
 &+& \log{\left( 1 +  \frac{\eta_{\rm 12}^{12}}{\eta_{\rm 12}^{16}}\frac{M_{\rm C,SN}}{M_{\rm O,SN}}\right)} +   \log{\frac{M_{\rm O,SN}}{M_{\rm Ni,SN}}}  
\end{eqnarray}

  Carbon-enhancement is
pronounced at shortest delays (i.e. lowest metallicity) reaching [C/Fe]$\sim 5$ as
can be seen in the bottom panels of Figure \ref{fig:CoverFe}.
These extreme values are caused by the fact that for extremely short delays,
Nickel is heavily depleted while spallation does not proceed onto the CO
layers thus preserving the original ($M_{\rm C,SN}=1M_{\odot}$) Carbon content.

 There  are  two classes of carbon stars in our model with a clear demarcation at or near $t_{\rm O}$ (see
 the corresponding [Fe/H] in Table 1).  
 These two separate classes is reminiscent of the observed separation of CEMP stars
 into C-rich (i.e. $t_{\rm delay}< t_{\rm O}$ in our model) and C-normal (i.e. $t_{\rm delay}> t_{\rm O}$ in our model) subgroups over the range $-4.0 < {\rm [Fe/H]} < -1.0$
(see Figure 3 in Aoki et al. 2007). Also, results from analysis of high-resolution spectra from the HERES survey (e.g, Barklem et al. 2005;  Lucatello et al. 2006) show  that at least 20\% of all stars at metallicity with [Fe/H] $< -2$ are enhanced in their [C/Fe] ratios by a factor of 10 or more above the solar value (i.e. [C/Fe]$ > 1$). Although the sample size is currently small, Beers \& Christlieb (2005) point out that 5 of 12, roughly 40\%, of stars known with  [Fe/H]$ < -3.5$, based on high-resolution spectroscopic studies, are strongly carbon enhanced. Below [Fe/H]$<-5$, the fraction is 100\% (2 of 2 stars  with [C/Fe]=4 for both HMP stars).

 The statistics mentioned above suggest that the strongly enhanced stars might have originated  from QN progenitors (i.e. the progenitors of the preceding SN)
near the upper mass limit of $40M_{\odot}$. These we expect to experience a QN with the shortest delays 
since the massive progenitors are expected to leave the  heaviest NSs with core densities closest to the instability (i.e. quark  deconfinement)
value. Using a Scalo IMF (Scalo 1986), we estimate about 1 strongly CEMP
star to form per 1000 C-normal star.
 Our model seems to capture these trends and   suggests that 
the fraction of CEMP stars tend to increase with decreasing metallicity.
 While one might look at the nature of CEMP stars as a matter of  the astrophysical origin of the carbon excess that is observed in these objects,
 we suggest that it is rather a  matter of Ni (and thus iron) destruction/depletion as suggested by our model.

 \subsection{Trends in Nitrogen}
 \label{sec:nitrogen}

\begin{figure*}[t!]
\begin{center}
\includegraphics[width=0.495\textwidth, height=0.45\textwidth]{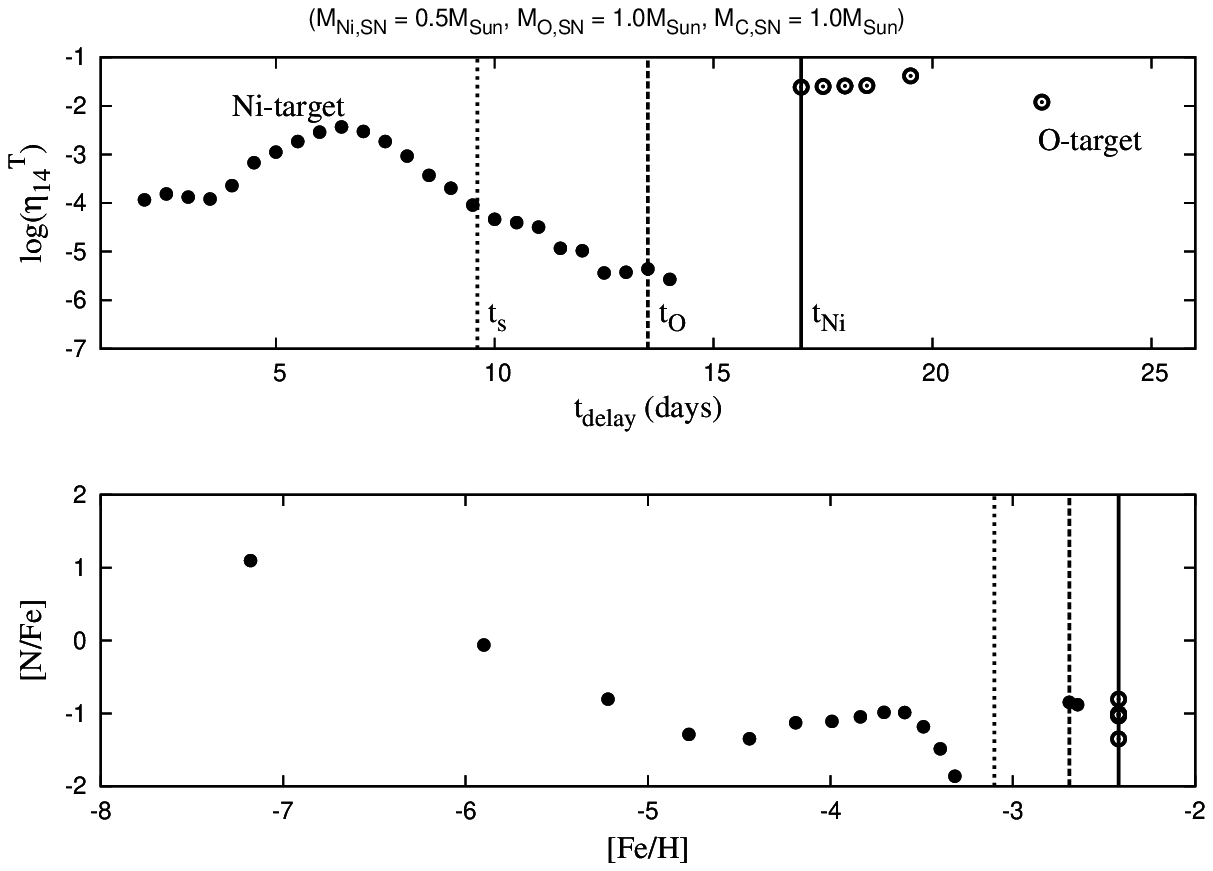}
\includegraphics[width=0.495\textwidth, height=0.45\textwidth]{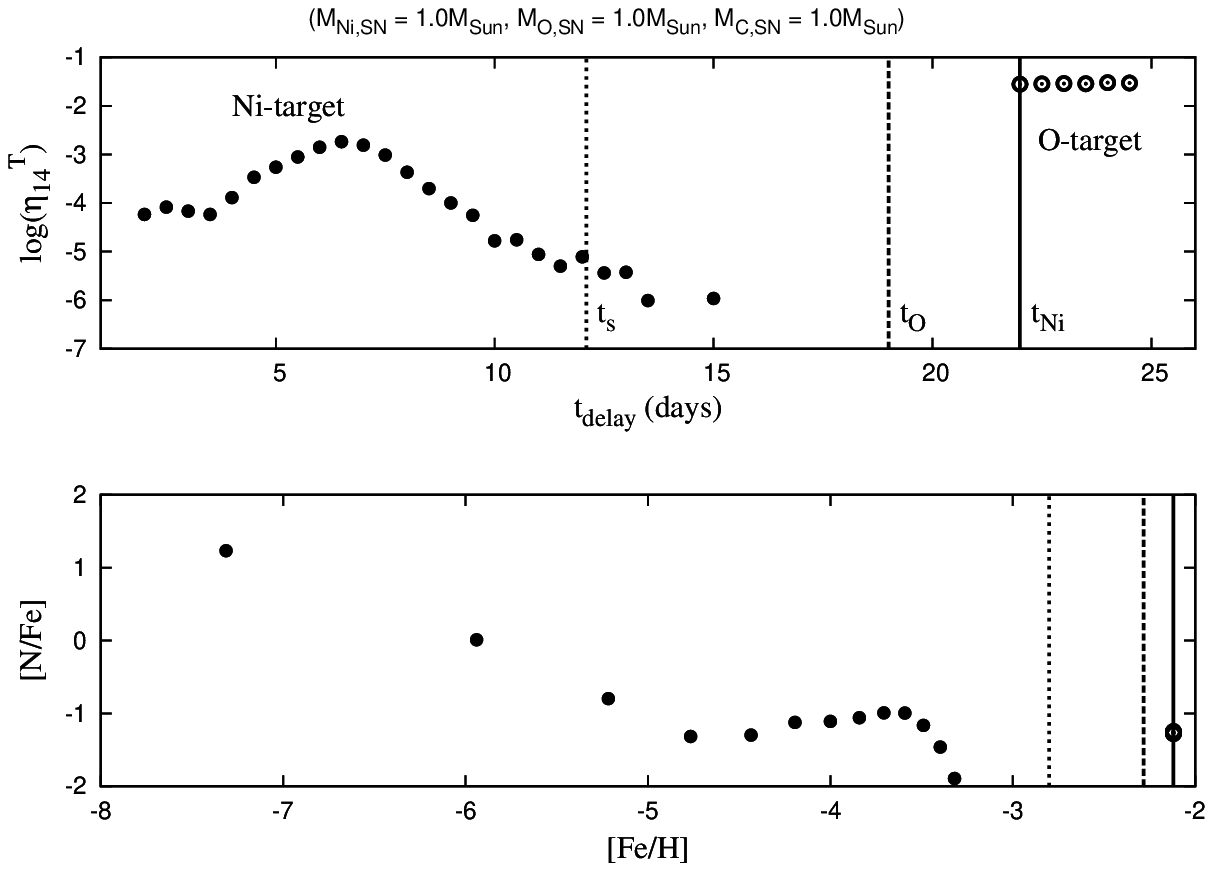}
\caption{{\bf Top panels}:  Nitrogen mass yields  versus
time delay ($t_{\rm delay}$) from Ni-spallation  ($\eta_{14}^{56}=M_{\rm N}/M_{\rm Ni,SN}$; solid circles)
 and  O-spallation   ($\eta_{12}^{16}= M_{\rm N}/M_{\rm O,SN}$; open circles).
{\bf Bottom Panels}: The corresponding [N/Fe] versus [Fe/H]. Left panels are  for  ($M_{\rm Ni,SN}=0.5 M_{\odot}, M_{\rm O,SN}=1.0 M_{\odot}, M_{\rm C,SN}=1.0 M_{\odot}$) while right panels are for 
($M_{\rm Ni,SN}=1.0 M_{\odot}, M_{\rm O,SN}=1.0 M_{\odot}, M_{\rm C,SN}=1.0 M_{\odot}$).
}
\label{fig:NoverFe}
\end{center}
\end{figure*}

  The dsQNe progenitors should contain a negligible amount of N
 compared to O and C since according to standard stellar models of 
 low metallicities massive stars, these should  not produce N  (e.g., Woosley et al. 2002).
   The top panels in Figure \ref{fig:NoverFe} compares N yield from Ni- and O-spallation. 
 The bottom panels shows the [N/Fe] versus [Fe/H]. 
 As expected for $t_{\rm delay} < t_{\rm O}$ (i.e. those with important Ni depletion) there is
little Nitrogen production in the outer shells. The only N produced
 was during  Ni-spallation but in negligible quantities compared to the N from O-spallation.  
 Nevertheless, for the $t_{\rm delay} < t_{\rm O}$ regime, [N/Fe] will increase
  with decreasing [Fe/H].

\begin{figure*}[t!]
\begin{center}
\includegraphics[width=0.495\textwidth, height=0.45\textwidth]{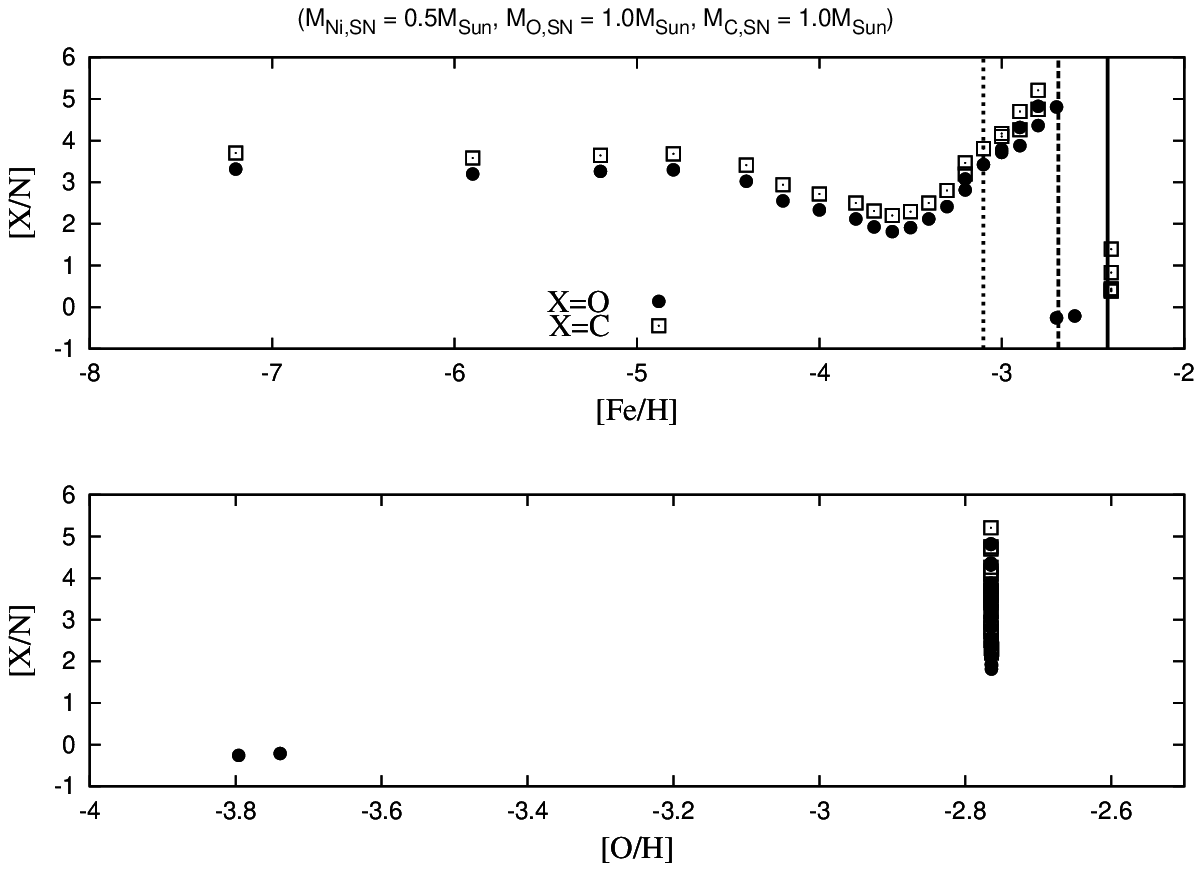}
\includegraphics[width=0.495\textwidth, height=0.45\textwidth]{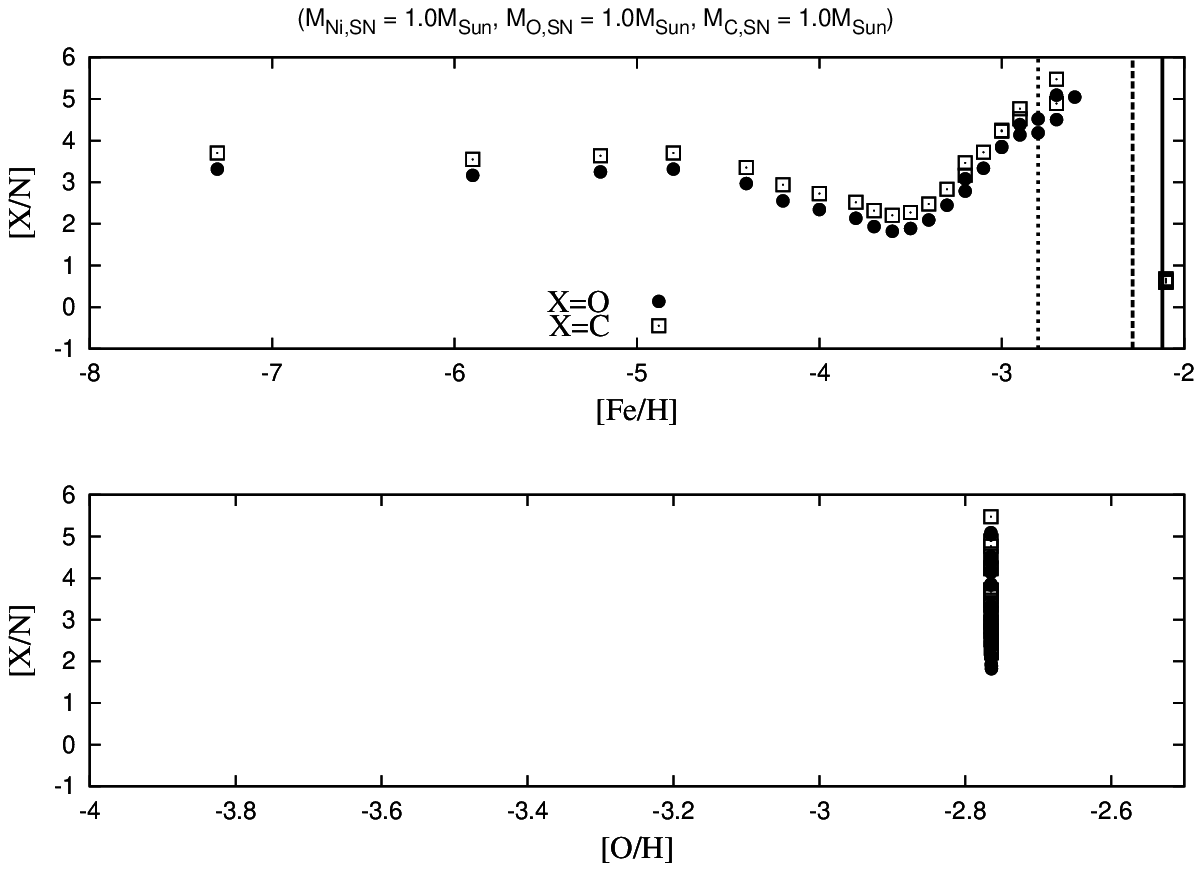}
\caption{{\bf Top Panels}: [O/N] (solid circles) and [C/N] (open squares) versus [Fe/H]. 
{\bf Bottom Panels}: [O/N] (solid circles) and [C/N] (open squares) versus [O/H].
 Left panels are  for  ($M_{\rm Ni,SN}=0.5 M_{\odot}, M_{\rm O,SN}=1.0 M_{\odot}, M_{\rm C,SN}=1.0 M_{\odot}$) while right panels are for 
($M_{\rm Ni,SN}=1.0 M_{\odot}, M_{\rm O,SN}=1.0 M_{\odot}, M_{\rm C,SN}=1.0 M_{\odot}$).}
\label{fig:CNOoverFe}
\end{center}
\end{figure*}

   Figure \ref{fig:CNOoverFe} shows [C/N] and [O/N] versus [Fe/H] (top panels) and
    versus [O/H] (bottom panels).
   As can be seen in the top panels, N is on average  less
 enhanced than O and C. This is because nitrogen is produced in small quantities
  during the inner-shell spallation and only when spallation proceeds to the outer layers (i.e.
  for $t_{\rm delay} > t_{\rm O}$) would N from O spallation becomes relatively more abundant. 
  We note that [O/H] is constant for $t_{\rm delay} < t_{\rm O}$ since 
   O in the outer layers exceeds by far the tiny amount generated from Ni-spallation.
   This explains the roughly constant [O/H] in the bottom panels. 
      For $t_{\rm delay} < t_{\rm O}$ we get  (see also the appendix)
 \begin{equation}
 \left[\frac{O}{H}\right]   \simeq  \log{\frac{N_{\rm O}}{N_{\rm H}}} - \log{\frac{N_{\rm O}}{N_{\rm H}}}\vert_{\odot} 
\sim    \log{\frac{(M_{\rm O, SN}/ 1  M_{\odot})}{(M_{\rm sw}/10^5M_{\odot})}} -2.78\ .
\end{equation} 
   However when spallation
   occurs in the outer shells (i.e. for $t_{\rm delay} > t_{\rm O}$), O is depleted
   which explains the two points at [O/H]$\sim -3.8$  in the bottom
   left panel of  figure \ref{fig:CNOoverFe}.
    For $M_{\rm Ni,SN}=1.0M_{\odot}$ (the bottom right panel), the neutron energy
    after Ni-spallation is much too low to spallate O. A range in $M_{\rm O, SN}$ or in the material swept up by the dsQN ejecta should
lead to a spread in   [O/H]  (see discussion in \S \ref{sec:plateau}).

   We close this section by briefly discussing our findings in the context 
   of observations and  models discussed in the literature:

   \begin{itemize}
   
   \item   Spite et al. (2005) reported high [N/O] ratios in a sample of EMP stars.  These
   are the Cayrel et al. (2004) sample that are not expected to have altered their surface values of CNO during their lifetimes (the ``unmixed stars"). 
   They concluded that it was likely that some source of primary N  was necessary to explain the observed high [N/O].
   Their arguments relies on the fact that 
   AGB stars had not had enough time to contribute to the ISM enrichment at such low metallicities and that massive stars are not producers of primary nitrogen (as predicted by standard stellar models, e.g., Woosley et al. 2002).   
    We argue that dsQNe with long delays (specifically those with $t_{\rm O} < t_{\rm delay} <  t_{\rm Ni}$), should lead
   to swept up clouds richer in N than those  swept-up and contaminated  by $t_{\rm delay} < t_{\rm O}$ dsQN explosions.
     We recall that for $t_{\rm delay}  > t_{\rm Ni}$,
    $\bar{\zeta}^{\rm O}\sim 4$ which means that it is mostly C ($A_{\rm P}= 16-4\sim 12$) that is produced.
    However, the $t_{\rm delay} < t_{\rm O}$ explosions will lead to  higher [N/Fe] values.  We speculate that  O-spallation by
    dsQNe in the  outer layers of the SN shells from the 20-40$M_{\odot}$ Pop. III stars
     as   the main source of   primordial N.

\item  It has been suggested in the literature (Ekstr\"om et al. 2008; Joggerst et al. 2010) that 
 replicating abundances of N in EMP and HMP stars may require rotationally induced mixing prior to the destruction of the progenitor star.
  The non-rotating Pop III model yields  did not reproduce the amount of N observed in HMP stars because it was not present in the initial pre-supernova progenitor models in sufficient quantities.  
      Chiappini et al. (2005) finds that 
 the measured [N/O]  can be explained by adopting the stellar yields obtained from stellar models which take into account rotation together with an extra production of N from massive stars born at metallicities Z below $10^{-5}$. This solution requires an
 increase of up to a few hundreds in N as compared to values expected from models  for rotating massive stars.
 In general, the yields from these models do not seem to produce  N and C abundances 
 self-consistently and require that Pop III stars were  rotating. Others argued that 
 the observed paucity of {\it very} N-rich stars  
   puts strong constraints on  the IMF at low metallicity (e.g. Komiya et al. 2007)
or  may require  a much more efficient dredge-up of carbon in
 low-mass, low-metallicity AGB stars  than shown by detailed evolution models available to date (see Pols et al. 2009; Izzard et al. 2009 for a discussion).
 We argue that dQNe model is a promising explanation for both the ubiquity of CEMP-s stars and the near-absence of NEMP stars
  without the need for a revision of dredge-up models of AGB stars. In  \S \ref{sec:IMF},
  we also discuss our model's implications to  the IMF of the first stars.
 
 \end{itemize}

\section{r-process and s-process elements}
\label{sec:sprocess}

 The total amount in mass of neutrons+protons formed following spallation
  is  $M_{\rm n+p}\simeq \zeta_{\rm net}M_{\rm QN}$. For a given $M_{\rm Ni,SN}$, dsQNe
  with $t_{\rm delay} < < t_{\rm Ni}$  lead to $M_{\rm n+p}\sim  0.1 M_{\odot}  \zeta_{\rm net, 100} M_{\rm QN, -3}$; each QN nucleon leads
  to an average of 100 spallated  neutrons+protons  with an average energy $E_{\rm n} \sim 10\  {\rm GeV}/\zeta_{\rm net}\sim 100$ MeV.  
    Neutron capture happens once the neutron energy is further reduced to $\le 30$ MeV 
    which means that s-processes in our model would occur preferably in short delays
    cases (i.e. most MP star) when $\zeta_{\rm net} >> 1$.  Efficient capture would occur 
     if neutron-capture timescale $\tau_{\rm cap.}=1/(n_{\rm A}\sigma_{\rm cap.} v_{\rm n})$
     is shorter than neutron lifetime, or $\tau_{\rm cap.} < 720$ s  which gives
    
    \begin{equation}
    \label{eq:tsprocess}
    t_{\rm delay} < t_{\rm s}\sim 9.6\ {\rm days}\   \left(\frac{M_{\rm Ni, SN}}{0.5M_{\odot}}\right)^{1/3}
    \frac{\sigma_{\rm cap., barn}^{1/3}E_{\rm n, 30}^{1/3}}{v_{\rm sn, 5000}}\ ,
    \end{equation}
       where we made use of $n_{\rm A} = M_{\rm A}/(4\pi R_{\rm in}^2\Delta R_{\rm in})$ and we assumed $\Delta R_{\rm in}\sim 0.1 R_{\rm in}$.
    The neutron energy is given in units of 30 MeV ($v_{\rm n}$ is the
    neutron thermal speed) and the neutron-capture cross-section, $\sigma_{\rm cap.}$,  in barns (Heil et al. 2006). 
     This is the third critical time in our model which is shown as the vertical dotted line in our figures.

     The weak dependence on the original (SN) Ni content implies that  $t_{\rm s} \sim 9.6\ {\rm days} /v_{\rm sn, 5000}$ 
    which corresponds to a universal value of  [Fe/H]$ < -4$ (from Table 1) in our model if the  velocity  of the preceding
     SN ejecta varies little from one progenitor to another.    A fraction of the neutrons will be captured
   before decaying into protons which should lead to  significant amounts of s-process elements.
   Even a slight neutron-capture of these thermal neutrons could
   lead to [Ba/Eu]$ > 1$ where we use Barium as a reference
   for s-process elements; as a comparison a pure r-process-enriched 
    star has [Ba/Eu] $\sim -0.8$ (e.g. Burris et al. 2000).

The QN ejecta  was shown to be an ideal site for r-processing of heavy elements (with $A>130$; see Jaikumar et al. 2007
for the QN as a novel r-process site). The elements   deposited into the SN shell
 can be quantified in terms of 
  [Eu/Fe]  (representing  the 3rd peak r-process elements) in our model as
\begin{eqnarray}
\left[\frac{Eu}{Fe}\right] &\simeq&  \log{\frac{N_{\rm Eu}}{N_{\rm Ni, SN}}} +   \log{\frac{N_{\rm Ni, SN}}{N_{\rm Fe}}} - \log{\frac{N_{\rm Eu}}{N_{\rm Fe}}}\vert_{\odot} \\\nonumber
&\simeq& -\log{\eta_{56}^{56}} +  \log{\frac{M_{\rm Eu}/10^{-8}M_{\odot}}{M_{\rm Ni, SN}/0.5 M_{\odot}}} -1.15
\end{eqnarray}
where $10^{-8}M_{\odot}$ is the average amount of Eu expected to form in a $10^{-3}M_{\odot}$
QN ejecta ($M_{\rm Eu}\sim 10^{-5}M_{\rm QN}$; Jaikumar et al. 2007).
Combining equation above with eq.(\ref{eq:FeoverH}) we get
  \begin{equation}
\left[\frac{Eu}{Fe}\right] \simeq -  [\frac{Fe}{H}] +  \log{\frac{M_{\rm Eu}/10^{-8}M_{\odot}}{M_{\rm sw}/10^5M_{\odot}}} - 3.57
\end{equation}
Equation above shows that [Eu/Fe] increases as [Fe/H]  decreases.
Since $ -7.2 < {\rm [Fe/H]} < -2$ then  $ -1.7 < [Eu/H] <  3.6$.\\

We close this section by discussing the implications of our findings above to CEMP stars:
  
  \begin{itemize}
  
  \item About 80\% of the CEMP stars have been shown to be enhanced in s-process elements (the so-called CEMP-s class; Beers\&Christlieb 2005).
  This can be accounted for in our model since s-process
 elements form preferably for short delays which is also the regime for CEMP stars.
 Recall (from \S \ref{sec:carbon})  that shorter delays preserve the original (SN) C while
 depleting Ni thus leading to high [C/Fe] values.

\item CEMP stars exhibit a wide variety of element abundance patterns, CEMP-s stars, CEMP-r stars and CEMP-r/s stars, which exhibit enhancements of the neutron-capture s-process elements, r-process elements and a combination of s- and r-process elements (Beers \& Chirstlieb 2005).    The QN ejecta is rich in r-process elements. If they survive the collision with the SN ejecta, these can be deposited in  the swept up gas/cloud thus leading
to CEMP-r stars. The lack of s-process in CEMP-r stars suggest that either
not enough neutrons are spallated or that most of them have decayed to
 protons before capture.  One can imagine a combination of the scenarios above leading to CEMP-r/s stars
  formed in clouds contaminated with s- and r-process elements.

  \item   It has been suggested since the majority of CEMP-s are known to be members of binary systems
 (e.g. Lucatello et al. 2005), these could have acquired their peculiar abundances from the companion.
 Enrichment by a companion AGB star would result in abundances of C, N, and O well above that with which the stars were initially born. 
 For example,  this has been proposed to explain  HE 1327$-$2326 (Aoki et al. 2006), although Frebel et al. (2008) present evidence that HE 1327$-$2326 is not a member of a binary system and thus could not have been enriched by an AGB companion.
Furthermore, in these accretion scenarios high level of C is often
     accompanied by  high levels of s-process abundances (e.g. Cohen et al. 2006) which is not always observed.
     In general it seems that  explaining CEMP peculiar composition with these models have proven difficult (e.g. McWilliam et al. 2009).    
      Our model provides a viable explanation of the variety of chemical abundances 
     observed in CEMP stars without appealing to mass transfer
     across a binary system.     
  
\end{itemize}

\begin{table}[t!]
\caption{Theoretical and numerical values of critical transition times (in days) in our model for different
 initial Nickel content in the SN envelope, prior to the QN explosion.
 The corresponding  [Fe/H] values (see Figure 1) are also shown.}
\begin{center}
\begin{tabular}{|c|c|c|c|}\hline
 $M_{\rm Ni,SN}/M_{\odot}$ &  0.1  &  0.5  & 1.0\\\hline\hline
 $t_{\rm s}$   &  5.6  &  9.6  & 12.1 \\\hline
 [Fe/H]$_{\rm s}$  & -3.8 &  -3.1 & -2.8 \\\hline\hline
  $t_{\rm O}$  theoretical (numerical) &  6.3 (6.3) &  14.0  (13.5) & 19.8 (19.1) \\\hline
  [Fe/H]$_{\rm O}$  & -3.6 &  -2.7 & -2.3 \\\hline\hline
   $t_{\rm Ni}$   theoretical (numerical) &  8.3 (8.3) &  18.6  (17.1) & 26.3 (22.1) \\\hline
   [Fe/H]$_{\rm Ni}$  & -3.1 &  -2.4 & -2.1 \\\hline
\end{tabular}
\end{center}
\label{default}
\end{table}%

 \section{Lithium production: $^7$Li and $^6$Li}
 \label{sec:spite}

 \subsection{$^7$Li}
    
   The Lithium spallated on a target $A_{\rm T}$ (Ni, O or C)  is given\footnote{In this paper, we ignore $^7$Li from BBN which should in principle be added to the final value obtained in our model.  Interestingly, it has been suggested that free thermal neutral injection on primordial nucleo-synthesis yield a successful solution for reducing the $^7$Li abundance (Albornoz 
V\'asquez et al. 2012). This idea warrants future investigation in the context of our model  given  the primary and secondary (in particular thermalized) neutrons that might escape the ejecta and stream ahead of dsQNe.} by (see appendix)
\begin{eqnarray}
\label{eq:ALI}
A (Li)  &=&  \log \frac{N_{^7{\rm Li}}}{N_{\rm A_{\rm T}}} + \log \frac{N_{A_{\rm T}}}{N_{\rm H}} +12   \\\nonumber
 &\simeq&  \log{\eta_{\rm 7}^{\rm A_{\rm T}}} + \log{\frac{M_{\rm A_T,SN}/1 M_{\odot}}{M_{\rm sw}/10^5M_{\odot}}}  + 6.23 \ .
\end{eqnarray}

Figure \ref{fig:Ni-plateau} compares lithium spallated from Ni  (solid circles) and from Oxygen (open circles).
The top panels show the mass-yields with the plateau from O-spallation clearly visible (see \S \ref{sec:plateaus} for
an explanation for the origin of the plateau in $\eta_7^{16}$).

 The corresponding $A(Li)$ given by equation above is shown in the
 middle panels with the lithium from O-spallation visible at the higher [Fe/H] values.
The  Ni-spallated lithium reaches higher values than the lithium from O-spallation but in our model
 there are reasons to believe that Li spallated in the inner layers
 will not survive: Li is easily destroyed in collisions with protons for temperature $T > 2.5\times 10^6$ K. 
For an adiabatic expansion of the preceding SN shell ($T_{\rm env.}\propto t^{-2}$), the $T_{\rm env.} < 2\times 10^6$ K yields
\begin{equation}
 \label{eq:tLi}
t_{\rm delay} > t_{\rm Li}\sim 10.5\ {\rm days}\ \frac{R_{\rm prog., 100}T_{\rm env., 10}^{1/2}}{v_{\rm sn, 5000}}\ ,
\end{equation}
where the initial radius of the progenitor, $R_{\rm prog., 100}$,  is in units of 100$R_{\odot}$ and the
initial temperature of the SN envelope, $T_{\rm env., 10}$, in units of $10^{10}$ K.
The above translates to (see Figure 1)
 \begin{equation}
\left[ \frac{Fe}{H}\right] > -3.0 \ .
 \end{equation}
  Lower [Fe/H] values can be achieved for explosions involving higher $v_{\rm sn}$ values. For example, taking $v_{\rm sn}= 10^4$  km s$^{-1}$  leads to $t_{\rm delay} > 5.25$ days or [Fe/H] $>$ -4 (see bottom panel in figure 1).
   In general  we expect $^7$Li   to be produced mainly in the outer Oxygen-rich layers in
     systems experiencing a QN with $t_{\rm delay}>  10$-11 days or, [Fe/H] $> -3.0$ for our fiducial values.
     However,  $^7$Li from inner-shell spallation may  survive in a narrow window in [Fe/H].
      The resulting swept-up cloud
     (and thus low-mass stars) in these rare events  will be imprinted with $^7$Li abundance
     above the observed Spite plateau.
  
       Figure \ref{fig:CO-plateau} compares lithium spallated from Oxygen to  that from Carbon for $t_{\rm delay} > t_{\rm O}$.
       This comparison applies to the case when the ($1M_{\odot}$) O-layer is replaced by  an ($1M_{\odot}$) C-layer.
     The top panels show the corresponding normalized mass-yields which shows a higher
     plateau in C target.  The high values of $\eta_7^{12}$ as compared to $\eta_7^{16}$ are understandable 
     since for the O-layer case most of the incident neutrons+protons are first used to 
     produce N and C before producing Li and other lighter elements.   
      The corresponding A(Li) values are shown
     in the middle panels with $3.3 < A(Li) < 3.5$ for C target and $2 < A(Li) < 2.4$ for O target.   However,  in a layered SN ejecta where the C layer
         overlays the O layer, the original (SN) C experiences much less spallation. In this case,
          Li originates mostly from the O-spallation with $2 < A(Li) < 2.4$.   We also add that Li from O-spallation 
          dominates when the C and O layers are mixed or when  C is less abundant than O.

\subsection{The plateau in A(Li)}
\label{sec:plateau}
     
       In \S \ref{sec:plateaus} we provide a natural  explanation for the  plateau in the relative 
       Li mass yield,  $\eta_7^{16}$.
       However A(Li) depends on  $M_{\rm O,SN}$ and $M_{\rm sw}$ which may vary from
        one source to another. 
     The observed 0.05 dex scatter in the A(Li) value from star-to-star would result from our model
      if  the ratio between the mass of  O in the SN ejecta,  $M_{\rm O,SN}$, and the mass of the swept-up pristine gas, $M_{\rm sw}$,
      vary little from one source to another.  From eq.(\ref{eq:ALI}), it can be seen
       that a scatter of no more than 10\% in $M_{\rm O,SN}/M_{\rm sw}$ can account for the 
          observed scatter.  We speculate on two possibilities:  
         (i)  It has been suggested that $M_{\rm sw}$ might be linearly dependent on the 
      explosion energy (e.g. Shigeyama\& Tsujimoto 1998). Thus if the mass
       of the SN ejecta (and thus  of  $M_{\rm O,SN}$) depend linearly on
        the SN energy, i.e. $M_{\rm SN} \propto E_{\rm SN}$ then equation (\ref{eq:ALI})
        suggests that an A(Li) plateau with a small scatter is not unrealistic in our model;   
     (ii)  The QN energy
           varies little  which should  give a narrow range in $M_{\rm sw}$; the QN explosion is triggered when the NS reaches quark
           deconfinement density in its core  which is a universal value. 
           Combined  with the fact that the narrow range in the dsQNe progenitor mass might 
             give a narrow range in $M_{\rm O,SN}$,  this could also lead to a narrow range in $M_{\rm O,SN}/M_{\rm sw}$ and  thus to a  small scatter.
     
\subsection{$^6$Li}

\begin{figure*}[t!]
\begin{center}
\includegraphics[width=0.495\textwidth, height=0.5\textwidth]{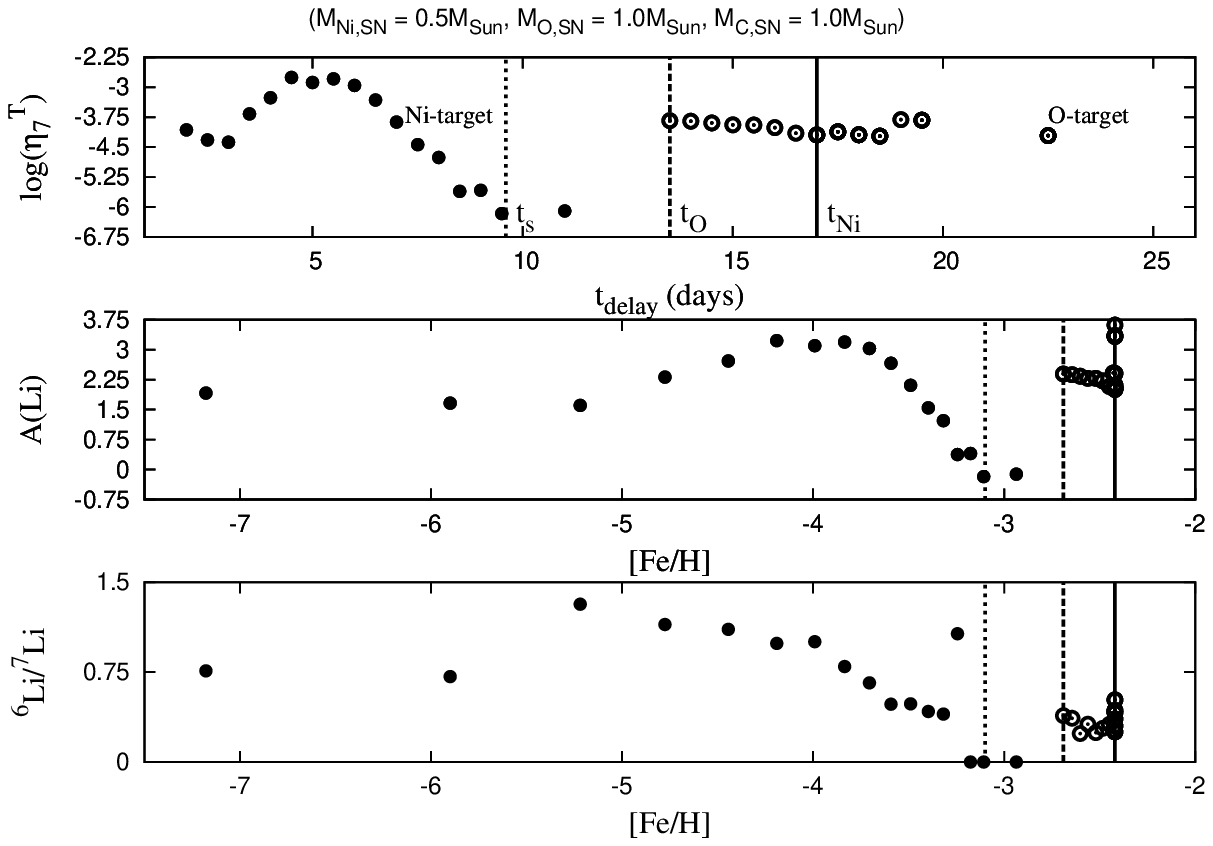}
\includegraphics[width=0.495\textwidth, height=0.5\textwidth]{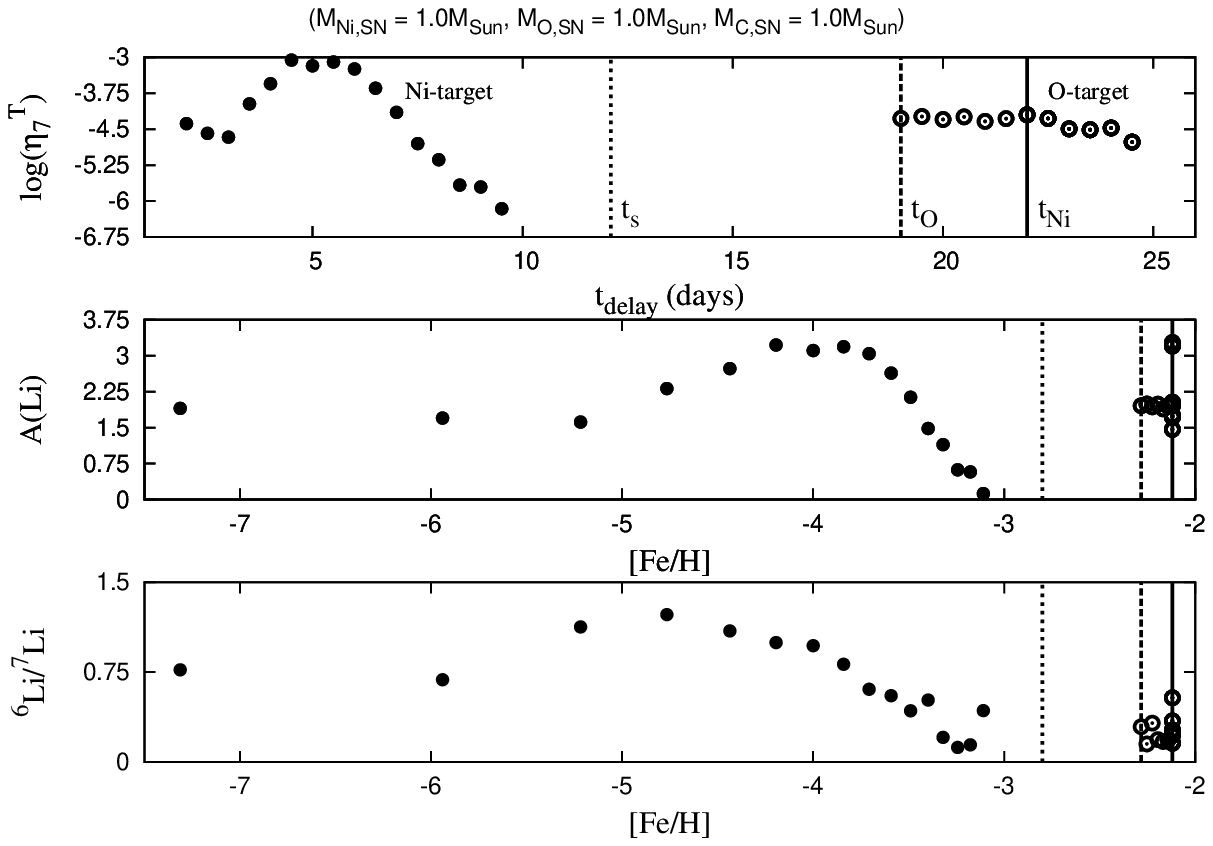}
\caption{{\bf Top panels}:   $^7$Li  mass yields  versus
time delay ($t_{\rm delay}$) from Ni-spallation  ($\eta_{7}^{56}=M_{\rm Li}/M_{\rm Ni,SN}$; solid circles)
 and  O-spallation   ($\eta_{7}^{16}= M_{\rm Li}/M_{\rm O,SN}$; open circles).
{\bf Middle Panels}: The corresponding A(Li) versus [Fe/H]. The plateau from O-spallation can be seen 
for $ t_{\rm O} < t_{\rm delay} < t_{\rm Ni}$. {\bf Bottom Panels}: The corresponding $^6$Li/$^7$Li ratio 
which also show a plateau for $ t_{\rm O} < t_{\rm delay} < t_{\rm Ni}$.  
Left panels are  for  ($M_{\rm Ni,SN}=0.5 M_{\odot}, M_{\rm O,SN}=1.0 M_{\odot}, M_{\rm C,SN}=1.0 M_{\odot}$) while right panels are for 
($M_{\rm Ni,SN}=1.0 M_{\odot}, M_{\rm O,SN}=1.0 M_{\odot}, M_{\rm C,SN}=1.0 M_{\odot}$).}
\label{fig:Ni-plateau}
\end{center}
\end{figure*}

\begin{figure*}[t!]
\begin{center}
\includegraphics[width=0.495\textwidth, height=0.5\textwidth]{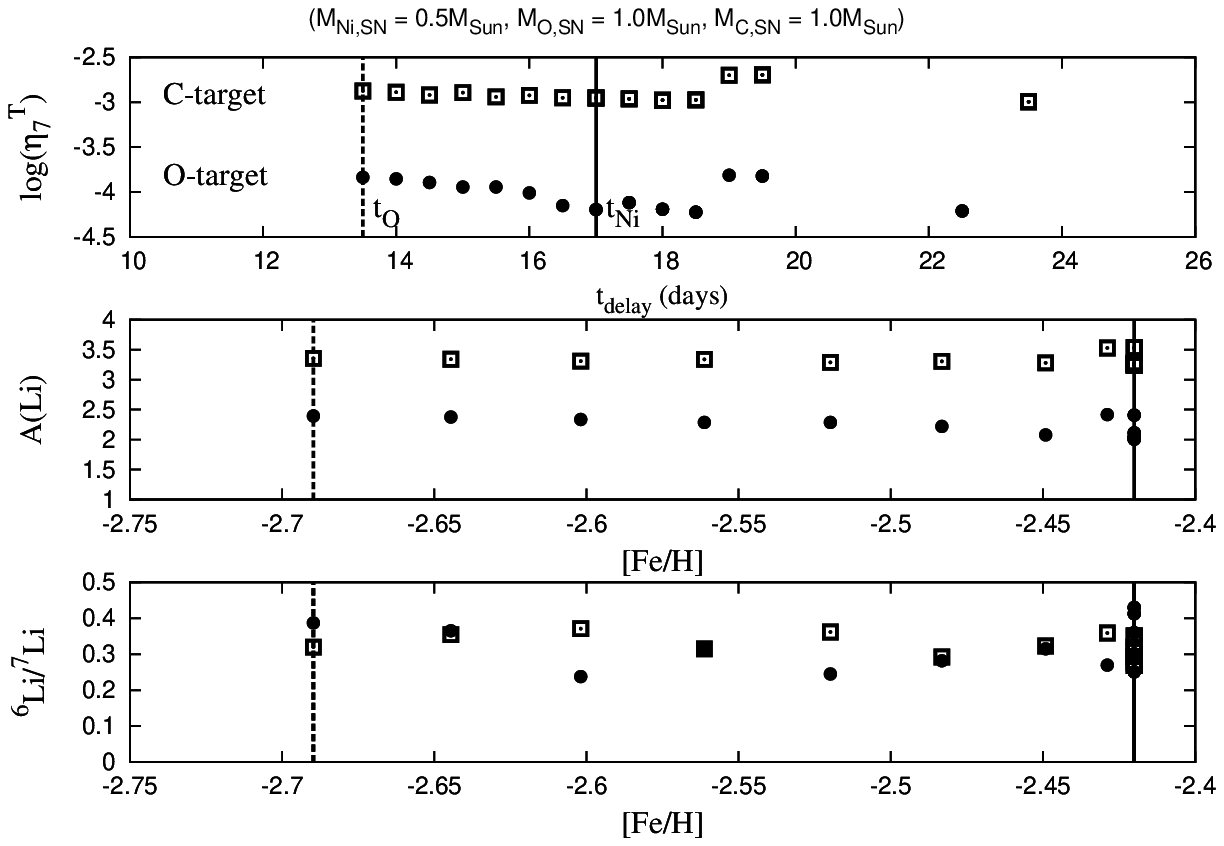}
\includegraphics[width=0.495\textwidth, height=0.5\textwidth]{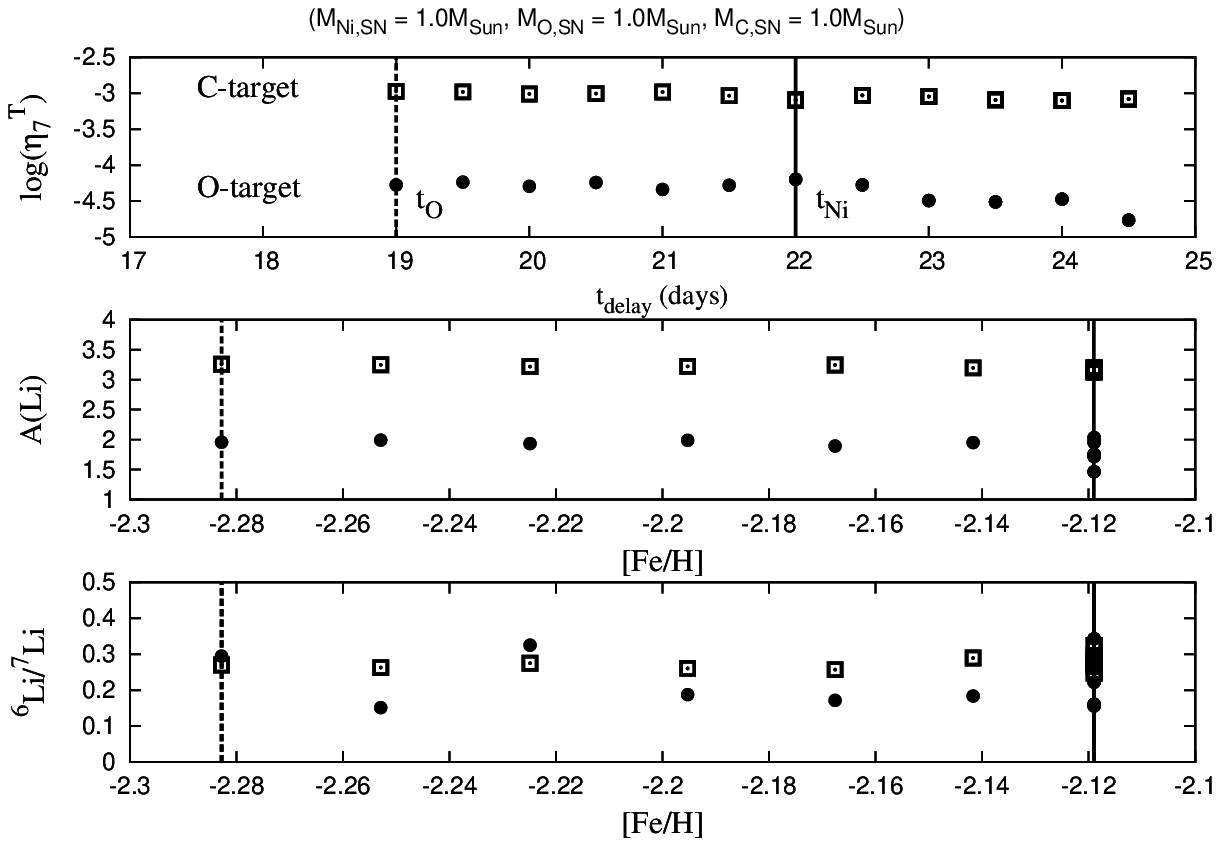}
\caption{{\bf Top panels}:  $^7$Li mass yields  versus
time delay ($t_{\rm delay}$) from O-spallation (solid circles; $\eta_{7}^{16}$)
 and  C-spallation   (open squares; $\eta_{7}^{12}$).
{\bf Middle Panels}: The corresponding A(Li) versus [Fe/H]. 
{\bf Bottom Panels}: The corresponding $^6$Li/$^7$Li ratio.  
Left panels are  for  ($M_{\rm Ni,SN}=0.5 M_{\odot}, M_{\rm O,SN}=1.0 M_{\odot}, M_{\rm C,SN}=1.0 M_{\odot}$) while right panels are for 
($M_{\rm Ni,SN}=1.0 M_{\odot}, M_{\rm O,SN}=1.0 M_{\odot}, M_{\rm C,SN}=1.0 M_{\odot}$).}
\label{fig:CO-plateau}
\end{center}
\end{figure*}

The bottom panels in figure \ref{fig:Ni-plateau} and figure  \ref{fig:CO-plateau} give
 an average $^6$Li/$^7$Li ratio
of the order of $\sim 0.3$.
Given  the spallation origin of Lithium in our model, it is not surprising that  $^6$Li is
   produced in amounts close to that of  $^7$Li.  
Since these two isotopes
  are formed in the same layer we expect temperature effects to further deplete 
   the $^6$Li abundance in our model to  $^6$Li/$^7$Li $ < 0.3$. 
 Our model is suggestive of  a plateau reminiscent of the one
  hinted at by observations ($^6$Li/$^7$Li$\sim 0.06\pm 0.03$; Asplund et al. 2006).     
 
  Since $^6$Li is  destroyed at  slightly smaller temperature (i.e. $T > 2 \times 10^6$ K)
 than $^7$Li, this means that $^6$Li would survive only in dsQNe with $t_{\rm delay} > t_{\rm ^6Li}\sim 11.5$ days for our fiducial values (see eq. \ref{eq:tLi}); i.e. for [Fe/H]$>$ -2.7 as shown in Figure 8.  This means as one approaches the  lower end/edge of  metallicity in the Spite plateau,  $^6$Li should start to decline slightly before that of   $^7$Li according to out model.   Lower [Fe/H] values can be achieved for explosions involving higher $v_{\rm sn}$ values (see eq. \ref{eq:tLi}).

To summarize this section, below we discuss  plausible implications of our model
 to the modelling of the convective envelope and atomic diffusion in MP stars:

\begin{itemize}

\item  In the traditional view, those Spite-plateau stars that have surface temperatures between 5700 and 6400 K have uniform abundances of $^7$Li because the shallow convective envelopes of these warm stars do not penetrate to depths where the temperature exceeds that for $^7$Li to be destroyed ($2.5\times  10^6$ K).  Stars to the left of the Spite plateau (cooler stars) posses  envelopes that  extend to such depths that their surfaces have lost $^7$Li to nuclear reactions.  In our model,  the stars to the left of the Spite plateau correspond to dsQNe with $t_{\rm delay} > t_{\rm Li}$ (i.e. $t_{\rm delay} > 10$-11 days or equivalently 
    [Fe/H] $> -3$ for our fiducial values).  Thus our model offers an explanation that is independent of the 
     question of the depth of the convective envelope in MP stars. Instead we speculate that
      the convective envelope might be  the same (i.e. operates similarly) in all low-mass halo stars and that  the cooler, low-mass ones 
      do not show Li simply because they were born in clouds swept-up and contaminated by 
       primordial dsQNe with $t_{\rm delay} < t_{\rm Li}$.

  \item    Assuming our model is a correct representation of the Spite plateau, it 
      could shed more light on the issue  of atomic diffusion
         in old stars. This would require a more quantitative and accurate estimate of   $A(Li)$,
          including contributions from both the O and C layers. 
     If it turns out that  true $A(Li)$ values  in our model  are  higher than  the Spite values then atomic diffusion as described in
       Richard et al. (2005) and Korn et al. (2006)  
       might be at play reducing the dsQNe values to the Spite values. However, values of $A(Li)$ close to 2
        would suggest otherwise.     
 \end{itemize}

\section{Model's implications and General Discussion}

   \subsection{Beryllium and Boron}

Spallation on Carbon targets yields
\begin{equation}
\bar{\zeta}\sim 
 \left\{ 
 \begin{array}{rl}
 1.5 &\mbox{if $ t_{\rm O} < t_{\rm delay}  < t_{\rm Ni}$\ since\ $E_1 \sim 0.76\ {\rm GeV}$,} \\
  2.8 &\mbox{if $ t_{\rm delay}  > t_{\rm Ni}$ since $E_1\sim E_0 = 10\ {\rm GeV}$,}
       \end{array}
 \right.
 \end{equation} 
which means that C-spallation would lead mainly to  Beryllium ($A_{\rm P}=12-2.8\sim 9$) and Boron
($A_{\rm P}= 12-1.5\sim 10$) formation.  Thus light elements such as Beryllium and Boron are simultaneously produced
 from spallation in the outer shells; these will be discussed in details elsewhere.  Let us simply note that
  our model predicts the existence of Beryllium-rich halo stars  for $t_{\rm delay} > t_{\rm Ni}$ reminiscent
   of the halo star HD 106038 (Smiljanic et al. 2008).  In fact, for $t_{\rm delay} > t_{\rm Ni}$,
   the $A>130$ elements processed in the QN ejecta (Jaikumar et al. 2007) should be preserved
    which should explain the Barium overabundance observed in HD 106038.
     Another possibility is that the light elements could be produced in HNe and in superbubbles (e.g.
     Parizot \& Drury 1999; Parizot 2000) if these are limited to no more than 3 per 100 SNe in order
     not to overproduce the light elements (See Fields et al. 2002).
      In our model,  Be abundance  increases with  $t_{\rm delay}$ (i.e. with Fe), a trend
   that should be testable in halo stars.  These two elements are fragile at temperatures $>3.5\times 10^6$ K
    which means they should be mostly associated with MP stars with $t_{\rm delay} > 10$ days (or equivalently, [Fe/H] $> -3.2$).
     Due to  the atomic and spectroscopic properties of Be and B, they have been less extensively observed.
      However, simultaneous observations of all three elements in MP stars should help  identify dsQNe  imprint.

   \subsection{Fluorine and Scandium}
   \label{sec:FandSc}
   
   Spallation in the inner Ni-layer produces     on average more Fluorine and Scandium than 
     the other sub-Fe elements  (see bottom panels in figures \ref{fig:XoverFe-in} and \ref{fig:XoverFe-out}).
     The single stable isotope of fluorine, $^{19}$F,  is not synthesized in the main nuclear reactions
        taking place in the cores of stars. The overabundance of this rare nuclide in model
         could become a key diagnostic of dsQNe in Galactic halo MP stars.
        In particular, since $^{19}$F is produced in higher quantities
        from Ni-spallation for short delays, we expect it to be even more prominent in CEMP
        stars.     Detection of overabundances of these
        two elements   in Galactic halo MP stars in general and in CEMP stars 
        in particular, could in principle be used to distinguish between  our model and other models.

\subsection{The primordial IMF and dsQNe}
\label{sec:IMF}

  The nature of the first stars, and hence the primordial initial mass function (IMF), is yet to be observationally constrained, but numerical models that proceed from well-posed cosmological initial conditions suggest that they are very massive, from $25$ to $500M_{\odot}$ (Bromm et al. 1999, 2002; Nakamura\&Umemura 2001; Abel et al. 2002;  Wise\&Abel 2007). Yet, 
  there seems to be no signatures of PISNe  in the
     Galactic Halo EMP stars  (e.g. Ballero et al. 2006; Cayret et al. 2004; Tumlinson et al. 2004).   It is possible that PISNe
      formed first and as the primordial gas became metal enriched, 
       a normal Pop. II stars started to form (e.g. Karlsson et al. 2008; see
       also Greif et al. 2010). In this case too it is puzzling that signatures    of PISNe have yet to be found.  
       
     Furthermore,  it seems that a successful reproduction of the general trends  in galactic halo stars may require
      a generation of stars that resemble massive stars of the 
 present-day IMF (i.e. from explosions resembling those of  normal  core collapse SNe). 
        Joggerst et al. (2010) found that the good match of IMF-weighted averages of SNe yields to  EMP star data suggests that SNe of $\sim 15 M_{\odot}$ with moderate explosion energies  were responsible for the bulk of early enrichment. 
      The fact that 15 $M_{\odot}$ progenitors must be included in the IMF average suggests that the metal-free stars that contributed the bulk of the metals to the early universe were of fairly low mass, extending down to the lower limit predicted for Pop III stars.
       Also, the ubiquity of CEMP-s stars and the near-absence of NEMP stars (see \S \ref{sec:nitrogen}) was
  also  used as a mean to constrain the IMF which lead to the suggestion that 
  an IMF biased toward intermediate-mass stars is required to reproduce the observed CEMP fraction in stars with metallicity [Fe/H] $<-2.5$
   (e.g. Izzard et al. 2009). However, these models also implicitly predict a large number of NEMP stars which is not seen.
 In summary, the best SN models  indicate progenitor masses of  $\sim 15$-$40M_{\odot}$ while
 currently, the primordial star formation scenario is favouring much higher initial masses for Pop III stars.

 Our model reconciles the two  if all galactic halo MP stars are formed
 in pristine clouds swept up by primordial dsQNe explosions. In a conventional or slightly top-heavy IMF,
      stars with $20 M_{\odot} < M_{\rm prog.} < 40 M_{\odot}$ would lead
       to dsQNe. The fragmentation of the  clouds swept up by primordial dsQNe 
       will lead to the first generation of low-mass stars with signatures reminiscent of EMP/UMP stars in the Galactic Halo.
        A daring conclusion is:  even if very massive Pop. III stars did form,
      as shown by simulations, these might not have been essential in explaining  
       the formation of low-mass galactic halo stars.  In Ouyed et al. (2009a), we  already argued that a normal 
       or slightly top-heavy  IMF of the oldest stars can be
     reconciled with a large optical depth as well as the mean metallicity  of the
     early intergalactic medium post re-ionization.

  \subsection{QNe as universal r-process site} 
  \label{sec:QNrprocess}
 
 Studies of r-process nucleosynthesis 
suggest that the site of formation of the heavy r-process elements ($A > 130$-140) must be associated with events that occur very early in the history of our Galaxy (see Truran et al. 2002 for a review). These studies show that  the r-process mechanism that operates in this environment must be robust, in the sense that even a single nucleosynthesis event seems capable of producing an r-process abundance distribution closely in agreement with that found in solar system matter.  
We have already discussed isolated QNe as plausible universal and efficient r-process site.  
The neutron-rich QN ejecta as it expands away from the neutron star was
shown to make mostly  $A>130$ elements  (Jaikumar et al. 2007). Here we showed that in dsQNe 
the neutron-rich QN ejecta collides with the preceding SN ejecta which
might lead -- besides spallation products -- to re-processing of the $A>130$ heavy elements  depending
on the strength of the interaction between the QN ejecta and the SN ejecta (i.e. $t_{\rm delay}$).
 In general it is not unrealistic  to assume that the resulting low-mass stars 
 should carry signatures of multiple r-process sites: (i) elements processed during the expansion of 
the QN ejecta; (ii) s-process elements from neutron-capture following spallation; 
(iii)  the r-process elements from the 
preceding SN; and (iv) any other processes in the SN shell
following reheating and shocking by the QN shock.

\subsection{QNe rate}
\label{sec:QNrate}

 Past estimate gives  the rate of QNe to be $< 1/100$ of core-collapse SNe (Staff
 et al. 2006; Jaikumar et al. 2007; Leahy\&Ouyed 2009; Ouyed et al. 2009). Out of
  these, 10\% are dsQNe. We thus expect low-mass MP stars in the Galactic halo to
  be extremely rare with an average occurrence   $< 1/ 1000$ of total stars.     
 In our model, the QN progenitors closer to the $40M_{\odot}$ limit would
 go off first. These will most likely lead to the most massive
 NSs and thus to  shortest time delays between the SN and the QN; thus the lowest [Fe/H] values.
Adopting a Scalo IMF (Scalo 1986), this means  that 
  the number of stars should decrease precipitously as one moves towards lowest [Fe/H]
   values. On the other hand,  the proportion of CEMP stars objects would show
   the opposite trend increasing dramatically towards lower, more negative, [Fe/H] values. 
   Unless dsQNe are more common than our current estimates (e.g. if QNe turn out to be more prominent in the early universe),  our model might be unable to account for the observed metal-poor stars with the lithium abundances at the level of the plateau. While we can argue for dsQNe as  a new mechanism to produce Lithium in the early universe, low dsQNe rate means that these might not necessarily play a relevant role in the Lithium plateau.

    \subsection{The expansion velocity $v_{\rm sn}$}
   \label{sec:vsndiscuss}
   
In our model, we assumed that the spallation reactions happened virtually instantaneously after the QN detonates. This made it possible to disregard the evolution of $v_{\rm sn}$ due to the dsQN (see \S \ref{sec:vsn}). However, the dsQN could modify $v_{\rm sn}$  by accelerating the ejecta. This would make $N_{\rm \lambda_{\rm sp}}$ (see eq. \ref{eq:Nsp}) decrease and therefore would decrease Fe destruction and enhance outer shell spallation. Before the dsQN, $v_{\rm sn}$ is more or less constant, because it takes a timescale in the order of years for the SN ejecta to reach the Sedov phase, whereas spallation is very inefficient (and  not likely to take place) in  dsQNe  with  $t_{\rm delay}$ that are larger than a few weeks.

\section{Conclusion}

We have shown that primordial dsQNe offer viable explanations for  the trends of MP Galactic halo stars, suggesting that these Halo stars (or at least a percentage of them)  might have formed from fragmentation of  the shells (i.e. pristine gas) swept-up by the 
  dsQNe occurring  in the wake of Pop. III stars.
   ``Iron  impoverishment"   by spallation of the inner Nickel layer of the SN material
  by the neutron-rich QN ejecta is at the heart of our model with the time delay ($t_{\rm delay}$)
  between the SN and QN explosions as one of the key parameters.  
  The generation of low-mass stars formed in the cloud swept-up by dsQNe
   are born with a range in metallicity  $-7.5 < {\rm [Fe/H]} < -2$ and even lower for time
   delays shorter than those considered in this work.
  We have shown that C and N trends observed in HMPs can be reproduced in 
  our model as well a neutron-capture elements from capture of the
 thermal (spallated) neutrons.  
 
    Primordial dsQNe offer a source of a post BBN, pre-galactic $^7$Li production
 (and other light elements)  with a  plateau  at 
     $ 2 < {\rm A(Li)} < 2.4$ from O-spallation  (and $3.3 < {\rm A(Li)} < 3.5$ from C-spallation;
      see \S \ref{sec:spite} why we favor the O-spallation origin of the plateau). We find that Li is formed in dsQNe
     with $t_{\rm delay} > t_{\rm Li}$ (with  $t_{\rm Li}\sim$ 10-11 days, i.e. [Fe/H] $> -3$, for our fiducial values) when spallation proceeds
     into the outer CO layers.      There are not enough results to draw the conclusion that primordial QNe could have a relevant role in the Lithium plateau; in particular the rate of dsQNe in the early universe might be too low to account for the observed metal-poor stars with the lithium abundances at the level of the plateau.  An investigation in a proper cosmological framework (e.g.  taking into account $^7$Li processed in BBN) is necessary before firm conclusions can be    reached.   At this stage we can only suggest  dsQNe as a plausible  new mechanism to produce Lithium (and other light elements) in the early universe.

  Given that the overall Universe was largely devoid of metals at the earliest times, 
  it is generally assumed  that low metallicity indicates old age. In this traditional view, the most MP stars are looked at as a ``probe"
 of the undiluted imprint of the first stars and, 
the lower the metal content of a star, the fewer
 instances of nucleosynthesis and recycling tat preceded its formation.
 Our model  suggests that the metal content of MP Galactic halo stars
 might not be related to age but rather to spallation from primordial dsQNe which instead depends
 on the properties of (i.e. mass of and Ni processed by)  the progenitor stars with $20M_{\odot} < M_{\rm prog.} < 40M_{\odot}$.
      The dsQN scenario, at least qualitatively, is able  to explain some features/abundances observed in
   old stars within a single event. More work, however, is still needed to show whether the scenario 
   still holds quantitatively.

\section*{Acknowledgements}
The research  of R. O. is supported by an operating grant from the
National Science and Engineering Research Council of Canada (NSERC).

\appendix

\begin{onecolumn}

\section{[X/Fe] and [X/O] in our model}

\subsection{Inner layers; $t_{\rm delay} < t_{\rm Ni}$}

Defining $N_{\rm X}^{\rm Ni}/N_{\rm X}^{\rm O}/N_{\rm X}^{\rm C}$ as the number of atoms of elements $X$ produced 
following spallation on Ni/O/C  targets, we have for  a
 product of atomic mass $A_{\rm X} < 56$
\begin{eqnarray}
\left[\frac{X}{Fe}\right] &=& \log{\frac{N_{\rm X}^{\rm Ni} +  N_{\rm X}^{\rm O} + \alpha N_{\rm X}^{\rm C} }{N_{\rm Fe}}} -  \log{\frac{N_{\rm X}}{N_{\rm Fe}}}\vert_{\odot}
= \left[\frac{X}{Fe}\right]_{\rm Ni} + \log{\left( 1 + \frac{N_{\rm X}^{\rm O} }{N_{\rm X}^{\rm Ni}} + \alpha \frac{N_{\rm X}^{\rm C} }{N_{\rm X}^{\rm Ni}}\right)}\\\nonumber
&=& \left[\frac{X}{Fe}\right]_{\rm Ni} + \log{\left( 1 +  \frac{\eta_{\rm X}^{16}}{\eta_{\rm X}^{56}}\frac{M_{\rm O,SN}}{M_{\rm Ni,SN}} + \alpha \frac{\eta_{\rm X}^{\rm 12}}{\eta_{\rm X}^{56}}\frac{M_{\rm C,SN}}{M_{\rm Ni,SN}} \right)} \ ,
\end{eqnarray}
where
\begin{equation}
\left[\frac{X}{Fe}\right]_{\rm Ni} = \log{\frac{N_{\rm X}^{\rm Ni}}{N_{\rm Fe}}} -  \log{\frac{N_{\rm X}}{N_{\rm Fe}}}\vert_{\odot}= \log{\frac{N_{\rm X}^{\rm Ni}}{N_{\rm Ni,SN}}} +  \log{\frac{N_{\rm Ni,SN}}{N_{\rm Fe}^{\rm Ni}} }  -  \log{\frac{N_{\rm X}}{N_{\rm Fe}}}\vert_{\odot}
 = \log{\frac{\eta_{\rm X}^{56}}{\eta_{\rm 56}^{56}}} + \log{\frac{56}{A_{\rm X}}} - \log{\frac{N_{\rm X}}{N_{\rm Fe}}}\vert_{\odot}\ .
\end{equation}
 Here $\alpha = (1 - \delta_{\rm XO})$ and $ \delta_{\rm XO}$ is the kronecker's symbol
  with $\delta_{\rm OO} =1$ when dealing specifically with oxygen. Naturally for $A_{\rm X} > 16$, [X/Fe] = [X/Fe]$^{\rm Ni}$ since
  we assume that no $A_{\rm X} > 16$ elements were present in the Nickel layer (i.e. none were processed in the inner SN envelope) prior to the QN
  explosion.
  Also, for $t_{\rm delay} < t_{\rm O}$, spallation in the outer layers is minimal
with   $\eta_{\rm X}^{16} = 1$ and $\eta_{\rm X}^{12} = 1$ in the equations above.   Solar abundances, $\log{\frac{N_{\rm X}}{N_{\rm Fe}}}\vert_{\odot} = A({\rm X})\vert_{\odot} - A({\rm Fe})\vert_{\odot}$ are taken from Asplund et al. (2009; see their Table 1).

\subsection{Outer layers; $t_{\rm delay} > t_{\rm Ni}$}

For $t_{\rm delay} > t_{\rm Ni}$, spallation in the Ni layer is minimal (or $M_{\rm Fe}\sim M_{\rm Ni,SN}$) which means
\begin{eqnarray}
\left[\frac{X}{Fe}\right]_{\rm O} &=& \log{\frac{N_{\rm X}^{\rm O} + \alpha N_{\rm X}^{\rm C} }{N_{\rm Ni,SN}}} -  \log{\frac{N_{\rm X}}{N_{\rm Fe}}}\vert_{\odot}
= \log{\frac{N_{\rm X}^{\rm O} + \alpha N_{\rm X}^{\rm C} }{N_{\rm O,SN}}} + \log{\frac{N_{\rm O,SN}}{N_{\rm Ni,SN}}}-  \log{\frac{N_{\rm X}}{N_{\rm Fe}}}\vert_{\odot}\\\nonumber
&=& \log{\eta_{\rm X}^{16}} + \log{\left( 1 + \alpha \frac{\eta_{\rm X}^{\rm 12}}{\eta_{\rm X}^{16}}\frac{M_{\rm C,SN}}{M_{\rm O,SN}} \right)}+ \log{\frac{56}{A_{\rm X}}} + \log{\frac{M_{\rm O,SN}}{M_{\rm Ni,SN}}}- \log{\frac{N_{\rm X}}{N_{\rm Fe}}}\vert_{\odot}  \ ,
\end{eqnarray}

 The [X/O] abundance is found from,
\begin{equation}
\left[\frac{X}{O}\right]  = \left[\frac{X}{Fe}\right] - \left[\frac{O}{Fe}\right] 
\end{equation}
where (for $t_{\rm delay} < t_{\rm Ni}$)
\begin{equation}
\left[\frac{O}{Fe}\right]_{\rm Ni} \simeq  \log{\frac{\eta_{\rm 16}^{56}}{\eta_{\rm 56}^{56}}} + \log{\left( 1 +  \frac{\eta_{\rm 16}^{16}}{\eta_{\rm 16}^{56}}\frac{M_{\rm O,SN}}{M_{\rm Ni,SN}} \right)} - 0.65
\end{equation}
and (for $t_{\rm delay} > t_{\rm Ni}$)
\begin{equation}
\left[\frac{O}{Fe}\right]^{\rm O} \simeq \log{\eta_{\rm 16}^{16}} +  \log{\frac{M_{\rm O,SN}}{M_{\rm Ni,SN}}}- 0.65
\end{equation}

The  [X/H] abundance is found from,

\begin{equation}
\left[\frac{X}{H}\right]  = \left[\frac{X}{Fe}\right] + \left[\frac{Fe}{H}\right] 
\end{equation}
with [Fe/H] given by equation (\ref{eq:FeoverH}).
 For example in the case of oxygen (for $t_{\rm delay} < t_{\rm Ni}$)
 \begin{equation}
 \left[\frac{O}{H}\right]  \simeq  \log{\eta_{16}^{56}} +
  \log{\left( 1 +  \frac{\eta_{\rm 16}^{16}}{\eta_{\rm 16}^{56}}\frac{M_{\rm O,SN}}{M_{\rm Ni,SN}} \right)} + \log{\frac{(M_{\rm Ni, SN}/ 1  M_{\odot})}{(M_{\rm sw}/10^5M_{\odot})}} - 2.82
 \end{equation}
 and (for $t_{\rm delay} > t_{\rm Ni}$)
  \begin{equation}
 \left[\frac{O}{H}\right]  \simeq  \log{\eta_{16}^{16}} + \log{\frac{(M_{\rm O, SN}/ 1  M_{\odot})}{(M_{\rm sw}/10^5M_{\odot})}} - 2.82
 \end{equation}
 Finally, 
 \begin{equation}
 A(X) = \left[\frac{X}{H}\right] + A(X)\vert_{\odot} \ .
\end{equation} 

\end{onecolumn}

\label{lastpage}


\begin{thebibliography}{99}


\bibitem[]{}  Abel, T., G. L. Bryan, \& M. L. Norman, 2002, Science 295, 93

\bibitem[]{}  Albornoz V\'asquez, D., Belikov, A., Coc, A., Silk, J., \& Vangioni, E.\ 2012, arXiv:1208.0443

\bibitem[]{}   Aoki, W. et al. 2005, ApJ, 632, 611

\bibitem[]{}   Aoki, W., Frebel, A.,  Christlieb, N., et al.\ 2006, \apj, 639, 897

\bibitem[]{}   Aoki, W. et al. 2007, ApJ, 655, 492

\bibitem[]{}   Aoki, W., Beers, T.~C., 
Carollo, D., et al.\  2010, ``Proceedings of the 11th Symposium on Nuclei in the Cosmos. 19-23 July 2010. Heidelberg, Germany. Published online at Published online at http://pos.sissa.it/cgi-bin/reader/conf.cgi?confid=100

\bibitem[]{} Asplund, M.,  Lambert, D. L.,  Nissen, P.E., et al.\  2006, ApJ, 644, 229 

\bibitem[]{} Asplund, M., Grevesse, N., Sauval, A. J., \& Scott, P. 2009, ARA\&A, 47, 481

\bibitem[]{} Barklem, P.~S., Christlieb, N., Beers, T.~C., et al.\ 2005, \aap, 439, 129

\bibitem[]{} Baym, G., Pethick, C., \& Sutherland, P. 1971b, ApJ, 170, 299

\bibitem[]{}  Ballero, S. K., F. Matteucci \& C. Chiappini, 2006, New Astron. 11, 306.

\bibitem[]{}  Beers, T., Preston, G., \& Shectman, S. 1992, AJ, 103, 1987

\bibitem[]{}  Beers T. C., Christlieb N., 2005, Ann. Rev. A\&A, 43, 531

\bibitem[]{} Bodmer, A.R.  1971, \prd,  4, 1601 

\bibitem[]{}  Bonifacio, P. \& Molaro, P. 1997, MNRAS, 285, 847

\bibitem[]{}  Bonifacio, P., et al. 2002, A\&A, 390, 91

\bibitem[]{}  Bonifacio, P., Molaro, P., Sivarani, T., et al.\ 2007, \aap, 462, 851

\bibitem[]{}  Bromm, V., Coppi, P.~S., \& Larson, R.~B.\ 1999, \apj, 527, L5

\bibitem[]{}  Bromm, V., Coppi, P.~S.,  \& Larson, R.~B.\ 2002, \apj, 564, 23

\bibitem[]{}   Bromm, V., \& Loeb, A., 2003, Nature, 425, 812

\bibitem[]{} Burris, D. L. et al. 2000, ApJ, 544, 302

\bibitem[]{}  Cayrel, R.,  Depagne, E. Spite, M. et al., 2004, A\&A, 416, 1117

\bibitem[]{}  Cayrel, R., Steffen, M., Chand, H., et al.\ 2007, \aap, 473, L37 

\bibitem[]{}   Charbonnel, C., \& Primas, F. 2005, A\&A, 442, 961

\bibitem[]{}    Chiappini, C., Matteucci, F., \& Ballero, S.~K.\ 2005, \aap, 437, 429

\bibitem[]{}  Christlieb, N., Green, P.~J., Wisotzki, L., \& Reimers, D.\ 2001, \aap, 375, 366

\bibitem[]{}  Christlieb, N. et al. 2002, Nature, 419, 904 

\bibitem[]{} Coc, A., Vangioni-Flam, E., Descouvemont, P., Adahchour, A., \& Angulo, C.\ 2004, \apj, 600, 544

\bibitem[]{} Cohen, J. G. et al. 2006, AJ, 132, 137

\bibitem[]{} Cowan, J.~J., Roederer,  I.~U., Sneden, C., 
\& Lawler, J.~E.\ 2011, RR Lyrae Stars, Metal-Poor Stars, and the Galaxy, 223 

\bibitem[]{}  Ekstr\"om, S., Meynet, G., Chiappini, C., Hirschi, R., \& Maeder, A. 2008, A\&A, 489, 685

\bibitem[]{} Evoli, C., Salvadori, S. \&  Ferrara, A. 2008, MNRAS, 390, L14

\bibitem[]{} Fields, B. D. et al. 2002, ApJ, 581, 389

\bibitem[]{} Fields, B.~D.\ 2011, Annual  Review of Nuclear and Particle Science, 61, 47

\bibitem[]{}  Frebel, A., Aoki, W., Christlieb, N., et al. 2005 Nature 434, 871-873

\bibitem[]{}   Frebel, A., Collet, R.,  Eriksson, K., Christlieb, N., \& Aoki, W.\ 2008, \apj, 684, 588

\bibitem[]{}  Frebel, A., \& Norris, J.~E.\ 2011, arXiv:1102.1748, to appear in Vol. 5 of textbook "Planets, Stars and Stellar Systems", by Springer, in 2012

\bibitem[]{}  Galama, T. J., et al. 1998, Nature, 395, 670

\bibitem[]{}  Greif, T.~H., Glover, S.~C.~O., Bromm, V., \& Klessen, R.~S.\ 2010, \apj, 716, 510

\bibitem[]{}  Heger, A., \& Woosley, S. E. 2002, \apj, 567, 532

\bibitem[]{}  Heger, A., Fryer, C. L., Woosley, S. E., Langer, N., \& Hartmann, D. H. 2003, ApJ, 591, 288

\bibitem[]{}   Heger, A., \& Woosley, S.~E.\ 2010, \apj, 724, 341

\bibitem[]{}  Heil, M., K\"appeler, F. \& Uberseder, E. 2006, Mem. Ita. Astron. Soc. 77, 922 

\bibitem[]{}   Hirschi R., 2007, A\&A, 461, 571

\bibitem[]{}   Hobbs, L. M., \& Duncan, D. K. 1987, ApJ, 317, 796

\bibitem[]{}  Honda et al. 2004, ApJ, 607, 474

\bibitem[]{} Hwang, U., \& Laming, J.~M.\ 2012, \apj, 746, 130

\bibitem[]{}  Iocco, F.\ 2012, Memorie della  Societa Astronomica Italiana Supplementi, 22, 19

\bibitem[]{1970} Itoh, N. 1970, Prog. Theor. Phys., 44, 291 

\bibitem[]{}  Iwamoto, K., Nomoto, K., \& Mazzali, P. A. 2003, Lecture Notes in Physics, 598, 243
     
\bibitem[]{}  Iwamoto, N., Umeda, H., et al.   2005, Science, 309, 15

 \bibitem[]{}   Izzard, R.~G., Glebbeek, E., Stancliffe, R.~J., \& Pols, O.~R.\ 2009, \aap, 508, 1359 
 
 \bibitem[]{} Jaikumar, P., Meyer, B.~S., Otsuki, K., \& Ouyed, R.\ 2007, \aap, 471, 227
 
 \bibitem[]{}     Joggerst, C. C. et al. 2010, ApJ,  709, 11
 
 \bibitem[]{}    Karlsson, T., Johnson, J.~L., \& Bromm, V.\ 2008, \apj, 679, 6

\bibitem[]{}  Karlsson, T., Bromm, V., \& Bland-Hawthorn, J.\ 2011, arXiv:1101.4024

\bibitem[]{}  Ker{\"a}nen, P., Ouyed, R., \& Jaikumar, P.\ 2005, ApJ, 618, 485
 
\bibitem[]{} Kobayashi, C., Umeda, H. Nomoto, K. et al. 2006, \apj,  653, 1145

\bibitem[]{}  Komiya, Y., Suda, T., Minaguchi, H., Shigeyama, T., Aoki, W., \& Fujimoto, M. Y. 2007, ApJ, 658, 367

 \bibitem[]{}   Korn, A. J. et al. 2006, Nature, 442, 657
        
 \bibitem[]{} Korn, A.~J.\ 2008, 14th Cambridge  Workshop on Cool Stars, Stellar Systems, and the Sun, 384, 33
  
\bibitem[]{}   Lambert, D.~L.\ 2004,  the new cosmology: Conference on Strings and Cosmology; The Mitchell Symposium on Observational Cosmology. AIP Conference Proceedings, Volume 743, p. 206

 \bibitem[]{} Leahy, D., \& Ouyed, R.\ 2008, \mnras, 387, 1193 

  \bibitem[Limongi et al.(2003)]{Limongi03} Limongi, M., Chieffi, A. \& Bonifacio, P, 2003, \apj, 594,L123

\bibitem[]{} Lucatello, S., S. Tsangarides, T. C. Beers, E. Carretta, R. G. Gratton, \& S. G. Ryan, 2005, \apj, 625, 825

\bibitem[]{}  Machida, M.N., Tomisaka, K., et al. Nakamura, F., \& Fujimoto, M.Y. 2005, \apj, 622, 39
       
\bibitem[]{}  Mazzali, P. A., Iwamoto, L., \& Nomoto, K. 2000, ApJ, 545, 407

 \bibitem[]{}  McCray R., 1985, in Spectroscopy of Astrophysical Plasmas, edited by A. Delgarno \& D. Layzer, p. 270
 
\bibitem[]{}    McWilliam, A., J. D. Simon, and A. Frebel, 2009, in astro2010: The Astronomy and Astrophysics Decadal Survey, volume 2010
of Astronomy, p. 200

\bibitem[]{}   Molaro, P., Primas, F., \& Bonifacio, P. 1995, A\&A, 295, L47

  \bibitem[]{}  Mucciarelli, A., Salaris, M., \& Bonifacio, P.\ 2012, \mnras, 419, 2195
  
  \bibitem[]{}    Nakamura, F., \& Umemura, M.\ 2001, \apj, 548, 19
   
  \bibitem[]{}  Nakazato, K., Sumiyoshi, K., \& Yamada, S. 2008, Phys. Rev. D, 77, 103006
  
  \bibitem[]{}    Niebergal, B., Ouyed,  R., \& Jaikumar, P.\ 2010, \prc, 82, 062801
  
  \bibitem[]{}  Nishimura, T.,  Aikawa, M., Suda, T., \& Fujimoto, M.~Y.\ 2009, \pasj, 61, 909
    
\bibitem[]{}  Nomoto, K., Tominaga,  N., Umeda, H., Kobayashi, C.,  \& Maeda, K.\ 2006, Nuclear Physics A, 777, 424 

\bibitem[]{}   Norris, J. E. et al. 2007, ApJ, 670, 774

\bibitem[]{}  Omukai, K., Thuribe, T., Schnerder, R., \& Ferrara, A., 2005, \apj, 626, 627

\bibitem[]{2002} Ouyed, R., Dey, J., \& Dey, M. 2002, A\&A, 390, L39

\bibitem[]{2005} Ouyed, R., Rapp, R., \& Vogt, C. 2005, ApJ, 632, 1001

\bibitem[]{2009}  Ouyed, R., \& Leahy, D. 2009, ApJ, 696, 562

 \bibitem[]{}   Ouyed, R., Pudritz,  R.~E., \& Jaikumar, P.\ 2009a, \apj, 702, 1575

 \bibitem[]{}    Ouyed, R., Leahy, D.,  \& Jaikumar, P.\ 2009b, in Proceedings for "Compact stars in the QCD phase diagram II (CSQCD II)", May 20-24, 2009, KIAA at Peking University, Beijing- P. R. China [arXiv:0911.5424]

%\bibitem[]{2010}  Ouyed, R.  et al. 2010, submitted  to MNRAS [arXiv:1010.5530]  

\bibitem[]{2012}  Ouyed, R., Kostka, M., Koning, N., Leahy, D.~A., \& Steffen, W.\ 2012, \mnras, 423, 1652 [arXiv:1010.5530] 

\bibitem[]{2011a} Ouyed, R., Staff, J.,  \& Jaikumar, P.\ 2011a, ApJ, 729, 60

\bibitem[]{2011b}  Ouyed, R., Staff, J.,  \& Jaikumar, P.\ 2011b, \apj, 743, 116 

\bibitem[]{2011c}  Ouyed, R. et al.\ 2011, Physical Review Letters, 107, 151103

\bibitem[]{}   Ouyed, R., \& Staff, J.~E.\ 2011, Submitted to \apj [arXiv:1111.3053]

\bibitem[]{}  Ouyed, R., \& Leahy, D.\ 2012, submitted to NJP [arXiv:1202.2400]

\bibitem[]{}  Parizot, E. \& Drury, L. 1999, A\&A, 349, 673

\bibitem[]{}  Parizot, E.  2000, A\&A, 362, 786

\bibitem[]{}  Pols, O.~R., Izzard, R.~G., Glebbeek, E., \& Stancliffe, R.~J.\ 2009, Publications of the Astronomical Society of Australia, 26, 327

\bibitem[]{} Prantzos, N., 2006,  A\&A, 448, 665 

\bibitem[]{}  Rebolo, R., Beckman, J. E., \& Molaro, P. 1988, A\&A, 192, 192

\bibitem[]{}  Richard, O., Michaud,  G., \& Richer, J.\ 2005, \apj, 619, 538
 
 \bibitem[Ricotti et al.(2002)]{Ricotti02} Ricotti, M., Gnede, N.~Y., \& Shull, J.~M. 2002, \apj, 575, 49

\bibitem[]{} Rollinde, E., Vangioni, E. \& K.A.Olive, K. A. 2006,  ApJ, 651,658 

\bibitem[]{} Ryan, S. G., Beers, T. C., Deliyannis, C. P. \& Thorburn, J. A. 1996, ApJ, 458, 543

\bibitem[]{}  Ryan, S. G., Norris, J. E., \& Beers, T. C. 1999, ApJ, 523, 654

 \bibitem[]{} Scalo, J.~M.\ 1986, Fundamentals  of Cosmic Physics, 11, 1
 
 \bibitem[]{}  Shigeyama, T., \& Tsujimoto, T.\ 1998, \apj, 507, L135
 
\bibitem[]{}  Smiljanic, R., Pasquini, L., Primas, F., et al.\ 2008, \mnras, 385, L93

\bibitem[]{} Sneden, C., McWilliam, A., Preston, G. W., et al. 1996, ApJ, 467, 819

\bibitem[]{}  Sneden, C., Roederer, I., Cowan, J., \& Lawler, J.~E.\ 2010, Nuclei in the Cosmos.,
"Proceedings of the 11th Symposium on Nuclei in the Cosmos. 19-23 July 2010. Heidelberg, Germany. Published online at Published online at http://pos.sissa.it/cgi-bin/reader/conf.cgi?confid=100, id.74"

\bibitem[]{}  Spite, F. \& Spite, M. 1982, A\&A, 115, 357

\bibitem[]{} Spite, F., \& Spite, M. 1993, A\&A, 279, L9

\bibitem[]{} Spite, M., et al. 2005, A\&A, 430, 655

\bibitem[]{}   Spite, M., \& Spite, F.\ 2010, IAU Symposium, 268, 201

\bibitem[]{}   Staff, J.~E., Ouyed, R., \& Jaikumar, P.\ 2006, \apj, 645, L145

 \bibitem[]{} Suda, T., Aikawa, M., Machida, et al. 2004, \apj, 611, 476
 
 \bibitem[]{}  Terazawa, H., 1979, INS-Report-338, University of Tokyo
 
\bibitem[]{}   Thorburn, J. A. 1994, ApJ, 421, 318
  
\bibitem[]{}  Truran, J. W., Cowan, J. J., Pilachowski, C. A., \& Sneden, C. 2002, PASP, 114,
1293

\bibitem[]{}  Tumlinson, J., A. Venkatesan, \& J. M. Shull, 2004, \apj, 612, 602

\bibitem[]{}  Umeda, H., Nomoto, K., 
\& Nakamura, T.\ 2000, The First Stars: Proceedings of the MPA/ESO Workshop Held at Garching, Germany, 4-6 August 1999, ESO ASTROPHYSICS SYMPOSIA. ISBN 3-540-67222-2. Edited by A. Weiss, T.G. Abel, and V. Hill. Springer-Verlag, 2000, p. 150

\bibitem[]{}  Umeda, H., \& K. Nomoto, 2002, Astrophys. J. 565, 385

\bibitem[]{}  Umeda, H., \& Nomoto, K. 2003, Nature, 422, 871

\bibitem[]{}   Umeda H. \& Nomoto, K. 2008ApJ,  673, 1014
   
\bibitem[]{}  Vogt, C., Rapp, R., \& Ouyed, R. 2004,
Nuc. Phys. A, 735, 543

\bibitem[]{}    Weiss A., Cassisi S., Schlattl H., Salaris M., 2000, ApJ 533, 413

\bibitem[]{}    Wise, J.~H., \& Abel, T.\ 2007, \apj, 671, 1559

\bibitem[]{} Witten, E. 1984, \prd , 30,  272 

\bibitem[]{}   Woosley, S. E., Heger, A., \& Weaver, T. A. 2002, Rev.
Mod. Phys., 74, 1015

\bibitem[]{}  Yoshida, N., Oh, S. P., Kitayama, T., \& Hernquist, L., 2008, \apj, 663, 687
 
  
\end{thebibliography}
\end{document}